\documentclass[fleqn,usenatbib]{mnras}

\usepackage{newtxtext,newtxmath}

\usepackage{graphicx}	
\usepackage{amsmath}	
\usepackage{amssymb}	
\usepackage{adjustbox}
\usepackage{threeparttable}

\usepackage{tabto}
\usepackage{float}
\usepackage{color}
\usepackage{amssymb} 
\usepackage{amsfonts,amsmath,amscd} 
\usepackage[subnum]{cases}
\usepackage{mathtools}
\usepackage{verbatim}
\usepackage{t1enc}
\usepackage[T1]{fontenc}
\usepackage{ae,aecompl}
\usepackage{graphics}
\usepackage{adjustbox}
\usepackage{caption}
\usepackage{eufrak} 
\usepackage{euscript} 
\usepackage[normalem]{ulem} 
\usepackage{makeidx} 
\usepackage[usenames,dvipsnames]{pstricks}
\usepackage{epsfig}
\usepackage{pst-grad} 
\usepackage{pst-plot} 
\usepackage{pstcol,times,pstricks,pst-node,pst-coil} 
\usepackage{multicol} 
\usepackage{multirow} 
\usepackage{indentfirst} 
\usepackage{natbib} 
\usepackage{setspace} 
\usepackage{threeparttable}
\usepackage{longtable}
\usepackage{pdflscape}

\newcommand{\Msun}{{\rm M}_{\odot}}
\newcommand{\Rsun}{{\rm R}_{\odot}}
\newcommand{\Lsun}{{\rm L}_{\odot}}

\newcommand{\bse}{{\sc bse }}
\newcommand{\mocca}{{\sc mocca }}
\definecolor{smalt(darkpowderblue)}{rgb}{0.0, 0.2, 0.6}

\title[CVs in GCs: bright and faint populations]
{MOCCA-SURVEY database I. Accreting white dwarf binary systems in globular clusters -- IV. cataclysmic variables -- properties of bright and faint populations}
\author[D. Belloni et al.]
{Diogo Belloni$^{1,2}$\thanks{E-mail: diogo.belloni@inpe.br}, 
Mirek Giersz$^2$\thanks{E-mail: mig@camk.edu.pl}, 
Liliana E. Rivera Sandoval$^{3}$\thanks{E-mail: Liliana.Rivera@ttu.edu},
Abbas Askar$^{2,4}$
\newauthor and Pawe\l{} Cieciel\k{a}g$^{2}$ \\
$^{1}$ National Institute for Space Research, Av. dos Astronautas 1758, S\~ao Jos\'e dos Campos, Brazil\\
$^{2}$ Nicolaus Copernicus Astronomical Centre, Polish Academy of Sciences, ul. Bartycka 18, PL-00-716 Warsaw, Poland\\
$^{3}$ Department of Physics, Box 41051, Science Building, Texas Tech University, Lubbock, TX 79409-1051, USA\\
$^{4}$ Lund Observatory, Department of Astronomy, and Theoretical Physics, Lund University, Box 43, SE-221 00 Lund, Sweden}
\begin{document}

\date{Accepted 2018 November 09. Received 2018 November 09.; in original form 2018 April 16}

\pagerange{\pageref{firstpage}--\pageref{lastpage}} \pubyear{2018}

\maketitle

\label{firstpage}

\begin{abstract}
%
We investigate here populations of cataclysmic variables (CVs) 
in a set of 288 globular cluster (GC) models evolved with the \mocca code.
This is by far the largest sample of GC models ever analysed with respect to CVs.
Contrary to what has been argued for a long time, we found that dynamical destruction of primordial
CV progenitors is much stronger in GCs than dynamical formation of CVs,
and that dynamically formed CVs and CVs formed under no/weak influence of dynamics have similar white dwarf mass distributions.
In addition, we found that, on average, the detectable CV population is predominantly composed of CVs formed via typical common envelope phase (CEP)
($\gtrsim$ 70 per cent),
that only $\approx$ 2--4 per cent of all CVs in a GC is likely to be detectable, and that
core-collapsed models tend to have higher fractions of bright CVs than non-core-collapsed ones.
We also consistently show, for the first time, that the properties of bright and faint CVs
can be understood by means of the pre-CV and CV formation rates, their properties at their 
formation times and cluster half-mass relaxation times.
Finally, we show that models following the 
initial binary population proposed by Kroupa and set with low CEP efficiency better 
reproduce the observed amount of CVs and CV candidates in NGC 6397, NGC 6752 and 47 Tuc.
To progress with comparisons, the essential next step is to properly characterize the candidates as CVs (e.g. by obtaining orbital periods and mass ratios).
\end{abstract}


\section{Introduction}

Cataclysmic variables (CVs) are interacting
binaries composed of a white dwarf (WD) undergoing dynamically and thermally stable 
mass transfer from a low-mass companion, usually a main 
sequence (MS) star \citep[e.g.][]{Warner_1995_OK,Knigge_2011_OK}.
They are expected to exist in non-negligible numbers in globular clusters
(GCs), which are natural laboratories for testing theories of stellar 
dynamics and evolution 
\citep[e.g.][for a review on CVs in GCs]{Knigge_2012MMSAI}.

Due to the high stellar crowding in GCs and their intrinsic faintness, CVs are difficult 
to identify in such environments. Therefore, space telescopes with high spatial 
resolution and sensitivity such as the \textit{Hubble Space Telescope} (HST) and the \textit{Chandra X-ray Observatory} are required to detect them. Until now the best 
studied GCs with respect to CV populations are 
NGC 6397 \citep{Cohn_2010}, NGC 6752 \citep{Lugger_2017}, 
$\omega$ Cen \citep{Cool_2013} and 47 Tuc \citep{Rivera_2017}.
The identification of CVs in these GCs has been carried out by identifying HST
optical counterparts to Chandra X-ray sources. Usually these counterparts show an 
H$\alpha$ excess (suggesting the presence of an accretion disc), they are bluer 
than the MS stars and several also show photometric variability in different bands.

In the core-collapsed\footnote{Core collapse is a process in which
the GC core evolves by releasing potential energy to the outer parts (via two-body relaxation) 
and thus becoming hotter and more compact, due to its negative heat capacity.
The collapse halts when heat sources (primordial/dynamically formed binaries, intermediate-mass black holes, etc.)
add energy to the core.}
clusters NGC 6397 and NGC 6752, \citet{Cohn_2010} and \citet{Lugger_2017}
found the CVs to be divided into two populations, a 
bright and a faint one.
On their optical colour-magnitude diagrams (CMDs), {\it bright} CVs lie close to the MS and
{\it faint} CVs close to the WD cooling sequence, being 
R $\approx$ 21.5 mag the cut-off between both populations.
Interestingly, in the non-core-collapsed clusters 47 Tuc and $\omega$ Cen,
only one CV population is observed, and is mainly composed of faint CVs.

Another interesting distinction between bright and faint CVs
in core-collapsed GCs is related to the level of mass segregation, which is 
intrinsically connected with the GC relaxation time (proxy for the GC dynamical age) and the CV
masses.
For core-collapsed GCs, bright CVs are more
centrally concentrated than faint CVs, and faint CVs
have similar radial distribution to MS turn-off point (MSTO) stars.
This result indicates that bright CVs are more massive than faint ones.
Also, bright CVs have more massive donors since CVs close to the
MS have their fluxes clearly dominated by the donor, which is usually the case for CVs close or above the
period gap\footnote{The orbital period distribution of CVs shows two peculiarities, a period gap around 2.15 h--3.18 h \citep{Knigge_2006} and a minimum period of $\sim80$ min \citep{Gansicke_2009}.}. Faint CV fluxes are otherwise dominated by the WD and/or accretion disc,
which is associated with CVs close to the period minimum, i.e.
WZ Sge-type progenitors. Indeed, \citet{Cohn_2010} and \citet{Lugger_2017} inferred the total mass range for the bright CVs 
and faint CVs to be $1.5 \pm 0.2$ and $0.8 \pm 0.2$ M$_\odot$, respectively.
This is consistent with typical WD masses in CVs 
\citep[$\sim 0.6-1.0$ M$_\odot$,][]{Zorotovic_2011} and donor masses of  $\sim 0.6$ and $\lesssim 0.1$ M$_\odot$,
for bright and faint CVs, respectively.

The case of 47 Tuc is extremely interesting. Even though the CVs in this 
cluster are predominantly faint, all of them are more centrally 
concentrated than MSTO stars.
The optical fluxes are indicatives of the donor mass, as we already discussed,
which implies then that most CVs in 47 Tuc have
low-mass donors.
This brings then that the CV masses are dominantly the WD masses.
\citet{Rivera_2017} inferred CV masses of $1.4 \pm 0.2$ M$_\odot$ for both bright
and faint CVs in 47 Tuc, which is similar to what has been found for bright CVs in 
NGC 6397 and NGC 6752.
However, this implies that, for the faint CVs, the inferred WD masses in 47 Tuc 
are $\sim 1.2$ M$_\odot$, which is much higher than the standard WD mass in CVs.
These authors explain these high WD masses as the consequence of either a net 
mass growth (due to an interplay between accretion and nova outbursts) or 
to X-ray selection effects, since the larger the WD mass (and in turn the smaller the WD radius),
the deeper the potential well, and thus the greater the X-ray
emission \citep[e.g.][]{Aizu_1973}.

One fact to be considered here is that the relaxation times of core-collapsed
GCs are much shorter than those of non-core-collapsed ones, since they
evolved faster, by reaching the core collapse before the Hubble time (dynamically old) 
and non-core-collapsed GCs are still going towards core collapse (dynamically young).
In the particular case of NGC 6397 and NGC 6752, the half-light relaxation times ($T_{\rm rel}$)
are $\sim$ 0.4 and $\sim$ 0.74 Gyr, respectively
\citep[][2010 edition]{Harris_1996}. In contrast, $T_{\rm rel}$ for 47 Tuc
and $\omega$ Cen are $\sim$ 3.5 and $\sim$ 12.3 Gyr, respectively.
In this way, mass segregation occurs faster in core-collapsed GCs than
in non-core-collapsed ones.

In the three initial papers of this series \citep{Belloni_2016a,Belloni_2017a,Belloni_2017b}, 
we discussed 12 specific \mocca models with a focus on the properties of
their present-day CV populations, of the present-day CV progenitors, and 
how CV properties are affected by dynamics in dense environments. 
In this paper, we extend our analysis, by simulating 288 new GC models,
with updated stellar/binary evolution in order to investigate the statistical 
properties of GC CVs.
In other words, our aim here is to complement these works with the 
objective of checking whether previous results remain, on a statistical basis, 
and of explaining the above-mentioned observational properties.
Even though we compare our results to observations of CVs in GCs,
we do not aim to reproduce the observations of any of them in particular
but instead in a statistical way.

\section{Methodology}
\label{methodology}

\subsection{\mocca code}
\label{moccacode}

In order to simulate the GC models, we utilised the MOnte-Carlo Cluster
simulAtor ({\sc mocca}) code developed by \citet[][and references therein]{Giersz_2013},
which includes the {\sc fewbody} code \citep{Fregeau_2004} to perform
numerical integrations of three- or four-body gravitational interactions and
the Binary Stellar Evolution ({\sc bse})
code \citep{Hurley_2000,Hurley_2002}, with the
upgrades described in \citet{Belloni_2018b} and \citet{Giacobbo_2018},
to deal with star and binary evolution.
\mocca assumes a point-mass with total mass equal to the enclosed Galaxy
mass at the specified Galactocentric radius to model the Galactic potential,
and uses the description of escape processes in tidally limited
clusters follows the procedure derived by
\citet{Fukushige_2000}.
\mocca has been extensively tested against $N$-body codes
\citep[e.g.][]{Giersz_2008,Giersz_2013,Wang_2016,Madrid_2017} and
reproduces $N$-body results with good precision, not
only for the rate of cluster evolution and the cluster mass distribution,
but also for the detailed distributions of mass and binding energy of
binaries.

\subsection{\bse code}
\label{bsecode}

The most important part of the \mocca code with regards to CVs and related
objects is the \bse code \citep{Hurley_2000,Hurley_2002}.
To overcome the shortcomings of the original {\sc bse} code with
respect to CV evolution \citep[][see section 5.2]{Belloni_2017b}, 
\citet{Belloni_2018b} updated the code in order to include state-of-the-art 
prescriptions for CV evolution.
This upgraded version allows accurate modeling of 
interacting binaries in which degenerate objects are accreting 
from low-mass main-sequence donor stars. 
A summary of main changes is provided in what follows, but
more details can be found in \citet{Belloni_2018b}. 

The main upgrade of the code is a revision of the mass transfer 
rate equation which is now based on the model of \citet{Ritter_1988} 
and has been properly calibrated for CVs. 
We also added the radius increase/decrease of low-mass 
main-sequence donors that is expected when mass transfer is
turned on/off, which is fundamental to reproducing the observed 
orbital period gap in CV populations.
New options for systemic and consequential angular momentum loss (CAML) have been also incorporated. 
It now includes the magnetic braking prescription of \citet{Rappaport_1983}, with $\gamma=3$, and 
CAML prescriptions described in \citet{Schreiber_2016}. Both angular momentum loss (AML) mechanisms 
are believed to be crucial during CV evolution, being the CAML caused by mass transfer and mass loss
from the system due to nova eruptions.
Finally, different stability criteria for dynamical and thermal mass 
transfer from MS donors were implemented. New criteria depend on the adiabatic 
mass-radius exponent, the mass-radius exponent of the MS star and on the assumed
CAML \citep{Schreiber_2016}.

We also updated the code with respect to the evolution of massive stars,
as described in \citet{Giacobbo_2018}, since their fates are likely crucial
in GC evolution.
Stellar  winds  have  been  updated  based  on  the
equations described in \citet{Belczynski_2010} and
\citet{Chen_2015}, which
depends on metallicity. This approach
includes a treatment of stellar winds following \citet{Vink_2001}
and \citet{Vink_2005},  
which is adequate  for  O-type  and  Wolf-Rayet stars. 
In addition to the description in  \citet{Belczynski_2010}, mass loss treatment
takes into account the dependence on  the  electron-scattering  Eddington ratio
\citep{Grafener_2008,Vink_2011,Vink_2017}.
These authors also included new fitting formulas for the core radii, as described 
in \citet{Hall_2014}.
Moreover, they included in \bse new recipes for core-collapse  supernovae.  
In  particular,  they  implemented both the rapid and the delayed models for 
supernova explosion described in \citet{Fryer_2012}. 
Finally, it was added a formalism to account  for  pair-instability  and  
pulsational  pair-instability supernovae, following the prescription of \citet{Spera_2017}.

An additional update, described in \citet{Kiel_2008}, is connected
with the possibility of neutron star formation through electron-capture supernovae , which
is a low-energy supernova occurring when an ONeMg core collapses due to electron captures 
onto the nuclei $^{24}$Mg, $^{24}$Na and $^{20}$Ne \citep[e.g.][]{Miyaji_1980}. 
We allow electron capture supernova when the core of a giant has mass in the range 
of $1.6-2.25$ M$_\odot$. In this case, we assume no kick associated with
the neutron star formation.

The new \mocca version used here allows us to have more realistic cluster evolution 
and to infer more accurate CV properties from our analysis.

\section{Models}
\label{models}


In all models, we assume that all stars are on the zero-age MS
when the simulation begins and that any residual gas from the star
formation process has already been removed from the cluster. Additionally,
all models have low metallicity (0.001), are initially at virial equilibrium, 
and have neither rotation nor mass segregation. Moreover, all models are evolved
for 12 Gyr which is associated with the present-day in this investigation.
With respect to the density profile, all models follow a \citet{King_1966} model,
and we adopted two values for the King parameter W$_0$: 6 and 9.
Regarding the tidal radius, we assumed two values, namely 60 and 120 pc.
Finally, as a measure of the cluster concentration, we have three different 
half-mass radii: 1.2, 2.4 and 4.8 pc.


In this work, we adopted two initial binary populations (IBPs). 
The IBP is defined here as the set of initial binaries in a GC model 
following determined distributions for their parameters: semi-major axis,
eccentricity, masses, mass ratio, and period.
%
%
The first IBP, defined as the Kroupa IBP,
corresponds to models constructed based on the IBP 
derived by \citet{Kroupa_1995b,Kroupa_INITIAL} and \citet{Kroupa_2013} 
with the improvements described in \citet{Belloni_2017c}.
%
%
The other IBP, defined as Standard IBP, follows `standard'
distributions: 
(i) a uniform distribution for the mass ratio in the range $(0,1]$;
(ii) a log-uniform distribution for the semi-major axis in the range $[10^{-0.5},10^{4.5}]$ R$_\odot$;
(iii) a thermal distribution for the eccentricity in the range $[0,1]$.

For each IBP, we simulated models with three different numbers
of objects (single stars + binaries), namely $400$k, $700$k, and $1200$k.
Models following the Kroupa IBP have 95 per cent primordial binaries and 
masses of approximately $4.72 \times 10^5$, $8.26 \times 10^5$ and $1.42 \times 10^6$  ${\rm M_\odot}$.
Models following the Standard IBP have binary fraction of 10 per cent,
and masses of about $2.57 \times 10^5$, $4.50 \times 10^5$ and $7.73 \times 10^5$  ${\rm M_\odot}$.


In all simulations, we have used the \citet{Kroupa_2001} canonical initial mass function (IMF), 
with star masses in the range between $0.08~{\rm M_\odot}$ and $150~{\rm M_\odot}$
 \citep[][]{Weidner_2013}.
We emphasize that the IMF in all our simulations is preserved. This
is achieved by applying a similar procedure as described in section 6.3 of
\citet[][see also \citet{Oh_2015,Oh_2016}]{Belloni_2017c},
i.e. we first generate an array of all stars and after that we pair the stars in a way consistent with
the assumed mass ratio distribution, which is different in both IBPs.

\begin{table} 
\caption{Initial GC conditions and binary evolution parameters.
For all models, we adopted the metallicity $Z=0.001$,
the canonical \citet{Kroupa_2001} IMF (i.e. with two stellar segments),
with masses between 0.08 M$_\odot$ and 150 M$_\odot$,
and assumed that none of the recombination energy is used
to assist the expulsion of the envelope during CEP.
The first column presents either the initial cluster property or the binary
evolution parameter, while the second column exhibits the values
associated with them. Given the combination of all parameters, we have 
a total of 288 GC models. See Section \ref{models} for more details.}
\label{TabMODELS}
\begin{adjustbox}{max width=\linewidth}
\noindent
\begin{threeparttable}
\noindent
\begin{tabular}{l|l}
\hline\hline
Initial binary population				&  Kroupa, Standard		\\
Number of objects [$\times10^5$]			&  4, 7, 12		\\
Mass [$\times10^5$ M$_\odot$]				&  2.57, 4.50, 4.72, 7.73, 8.26, 14.2 	\\
Binary fraction						&  95 \% , 10  \% 			\\
King model parameter					&  6, 9				\\
Tidal radius [pc]					&  60, 120			\\
Half-mass radius [pc]					&  1.2, 2.4, 4.8		\\
Fallback						&  yes, no			\\
CEP efficiency 						&  0.25, 0.50, 1.00 		\\
\hline\hline
\end {tabular}
\end{threeparttable}
\end{adjustbox}
\end{table}


For each initial cluster configuration, we simulated models with three values
for the common envelope phase (CEP) efficiency, namely $\alpha=$ 0.25, 0.5 and 1.0.
In addition, we assumed that none of the recombination energy helps in the CE ejection and
that the binding energy parameter is  determined based on the giant
properties, as described in \citet[][see their appendix A]{Claeys_2014}.
The CAML prescription adopted here is the
one postulated by \citet{Schreiber_2016}, which is used to explain
several features associated with CV evolution.
Additionally, for massive stars, we assumed the delayed core-collapse
supernova model, which is described in \citet{Fryer_2012}. Supernova natal kicks
for neutron stars are distributed according to \citet{Hobbs_2005}.
In the case of black holes, they are also described according to \citet{Hobbs_2005}
or reduced according to mass fallback description given by \citet{Fryer_2012},
again for the delayed core-collapsed supernova model.
All other binary evolution parameters are set as in \citet{Hurley_2002}.

All the above-discussed variables in our modelling (i.e. binary evolution parameters, 
IBPs, and initial cluster conditions) are summarized in Table \ref{TabMODELS}.

\section{Cluster Evolution}
\label{clusterevol}

\begin{figure}
   \begin{center}
    \includegraphics[width=0.95\linewidth]{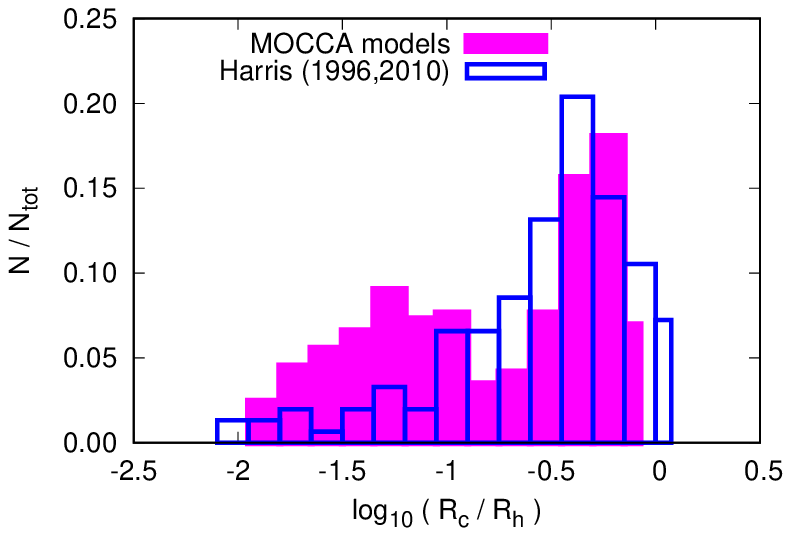}
    \includegraphics[width=0.95\linewidth]{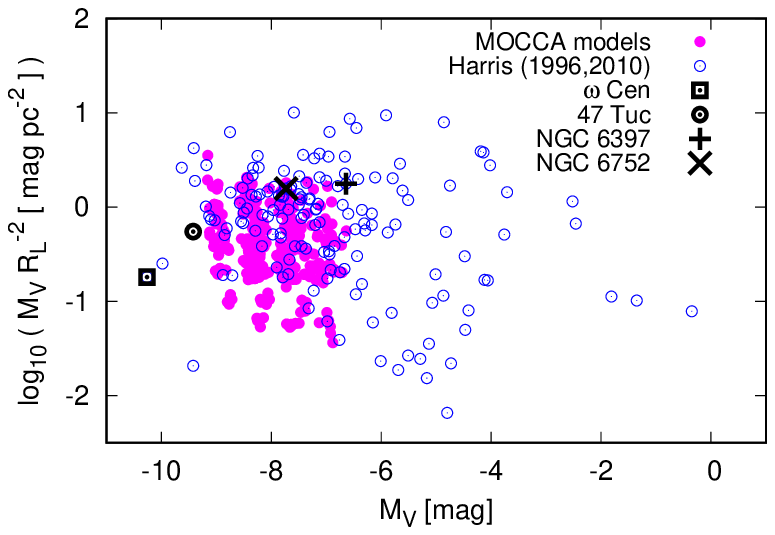}
    \includegraphics[width=0.95\linewidth]{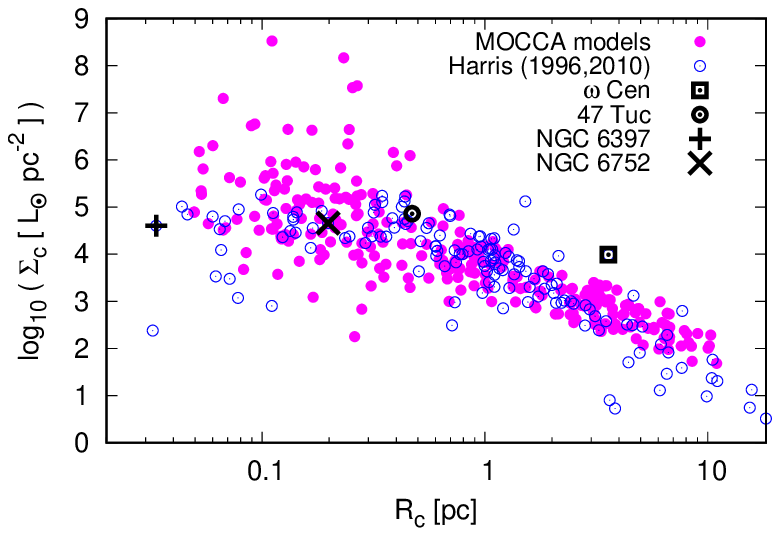}
    \end{center}
  \caption{Comparison between observational properties of 
  Galactic GCs (blue open points/histogram) and 
  present-day models (magenta filled points/histogram), 
  where four specific GCs are highlighted (black points). 
  In top panel, we compare the distribution of core to half-light radii.
  In the middle panel, the x-axis is the cluster V-band absolute 
  magnitude and y-axis is the average surface brightness inside 
  the half-light radius.
  In the bottom panel, we show the central surface brightness as a 
  function of the core radius.
  Observational data extracted from \citet[][ updated 2010]{Harris_1996}.}
  \label{Fig0}
\end{figure}


Among our 288 GC models, there are models which initially collapse due to mass segregation and this initial collapse is halted by binary energy generation or intermediate-mass black hole formation. Subsequently, during the long-term evolution of the GC models, the clusters can again evolve towards core collapse as they exhaust their initial population of binaries that were supporting the cluster evolution via binary burning during the balanced phase. Such GC models can undergo core collpase within a Hubble time and are characterized by smaller initial $T_{\rm rel}$.
Such core collapse is deeper and central density becomes large before sufficient number of binaries can dynamically form to prevent further collapse.

With respect to the binary fraction, we notice that for models
set with the Kroupa IBP (initially 95 per cent) it drops drastically 
during the early evolution. This is because most binaries in such
models are soft and the denser the cluster, the quicker their disruption.
In addition, while comparing predicted and observed binary
fractions in GCs, \citet{Leigh_2015} were able to conclude 
that only high initial binary fractions (set with the Kroupa
IBP, i.e. binaries with a significant fraction of very wide 
binaries) combined with high initial 
densities can reproduce the observed anti-correlation between 
the binary fraction (both inside and outside the half-mass radius)
and the total cluster mass \citep{Milone_2012}. 
Moreover, \citet{Belloni_2017c} further investigated the impact of
the Kroupa IBP in GCs. These authors compared predicted and observed 
properties of GC CMDs and managed to further improve properties of 
initial binaries in order to be incorporated into numerical 
simulation investigations.  They conclude that their modifications 
to the Kroupa IBP bring present-day GC models even closer to real GCs.

Finally, in order to check whether our models have present-day properties
consistent with the Galactic GC population, we show in Fig. \ref{Fig0} 
a comparison between our models and real GCs.
In the top panel, the distributions of core to 
half-light radii $(R_c/R_L)$ of models and real GCs are shown.
In the middle panel, we compare in the plane of
absolute magnitude ($M_V$) versus average surface brightness ($M_V R_L^{-2}$).
Finally, the central surface brightness ($\Sigma_c$) as a function of the 
core radius ($R_c$) is provided in the bottom panel.

Notice that our models lie in the region of massive and intermediate-mass
real GCs, and we only miss the low-mass GCs.
We also notice that there is a portion of our models in which values of the central surface brightness exceed those of real GCs $(\Sigma_c \gtrsim 10^6~\Lsun~{\rm pc}^{-2})$.
Such models have massive intermediate-mass black holes
that can have a deeper potential well and this contributes 
to the increase in central surface brightness. 
They also contribute to the excess seen in the distribution
of core to half-light radii for values in which 
$\log_{10}(R_c/R_L) < -0.8$. Below this value, $\sim90$ per cent
of the models host an intermediate-mass black hole.
More details about influence of black holes in shaping global 
GC properties can be found in \citet{Askar_2018b} and \citet{Askar_2018a}.

Such a comparison clearly shows that we have amongst our models cluster in
a reasonable range of concentrations, central surface brightness and relaxation times.
To sum up, we showed that our models are consistent with a reasonable part of real GCs.
We notice that such a conclusion is not surprising, given that our initial
models are similar to part of those shown in 
\citet{Askar_2016b}, \citet{Askar_2018b} and \citet{Askar_2018a}, and these
authors managed to show that their models are more or less representative of 
the Galactic GC population.

In order to properly compare our results with observations,
we now separate our models into two groups, 
which will correspond to our {\it core-collapsed} and 
{\it non-core-collapsed} models.
Since most core-collapsed real GCs have high central surface brightness
($\Sigma_c\gtrsim10^3$ $\Lsun$ pc$^{-2}$) and 
are very compact ($R_c\lesssim0.2~{\rm pc}$), we utilize such values to
define our core-collapsed and non-core-collapsed models.
The motivation for such a distinction is based on the
fact that the observational definition of core-collapsed and 
non-core-collapsed clusters can be ambiguous. That definition takes
into account only the current observational status of the cluster, and not
its whole evolution.
For instance, \citet{Heggie_2008} show that M4  
(which profile is a classic King profile typical of 
a non-core-collapsed cluster) is actually a post-collapse 
cluster, having its core sustained by binary burning. 
This means that considering the whole cluster evolution, M4 would be a core-collapsed cluster. 
But when only its current surface-brightness profile is considered, it is classified as non-core-collapsed. 
In a similar way, \citet{Giersz_2009} show that NGC 6397 
is also a post-collapse GC. This suggests that clusters currently classified as 
non-core-collapsed could have naturally undergone core-collapse 
at earlier times and vice versa. Such behaviour is 
related to the cluster gravothermal oscillations on time-scales of 
a few hundred million years.

After applying such criteria, we found that the fraction of models considered
core-collapsed is only $\sim30$ and $\sim20$ per cent, for those set with the
Kroupa and the Standard IBPs, respectively. 
This indicates that our sample of models are mainly composed of 
models that have properties much closer to non-core-collapsed
real GCs than core-collapsed ones.

\section{Destruction Rate of Primordial CV Progenitors}

We start our analysis quantifying the rate of destruction of
primordial CV progenitors with respect to the initial stellar
encounter rate, given by $\Gamma = \rho_0^2 r_c^3 \sigma_0^{-1}$
\citep{Pooley_2006},
where $\rho_0$, $r_c$ and $\sigma_0$ are the central density,
the core radius and the mass-weighted central velocity dispersion,
respectively. We note that $\Gamma$ can be interpreted as
an indicator of the strength of dynamics that one would expect
during the cluster evolution.

Before proceeding further, it is important to clearly define some
terms that will be used in what follows. We call {\it CV progenitors}
all binaries that somehow become CV and survive up to the present day.
Within the CV progenitors, those that are primordial binaries
are called {\it primordial CV progenitors}. From these definitions,
a primordial binary that becomes a CV via CEP, without having
its components altered via dynamics, is a primordial CV progenitor.
Alternatively, a primordial binary that has, for instance, one of its 
components replaced in a dynamical exchange interaction is just
a CV progenitor.
Finally, dynamically or thermally unstable CVs that do not survive 
up to the present day are not considered in this work.

\begin{figure}
   \begin{center}
    \includegraphics[width=0.95\linewidth]{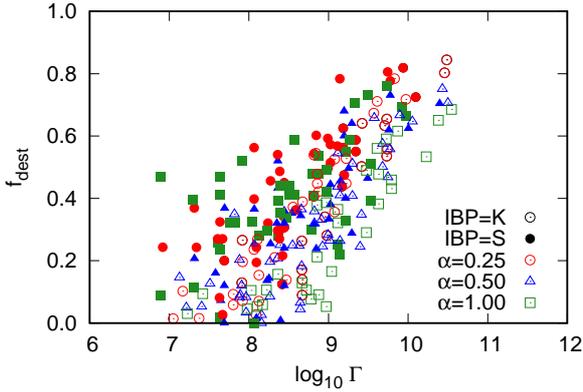}
    \end{center}
  \caption{Fraction of destroyed primordial CV progenitors 
  ($f_{\rm dest}$)
  versus the initial stellar encounter rate ($\Gamma$). 
  Note the clear correlation between $f_{\rm dest}$ and $\Gamma$.}
  \label{Fig1}
\end{figure}

\begin{figure*}
  \begin{center}
    \includegraphics[width=0.9\linewidth]{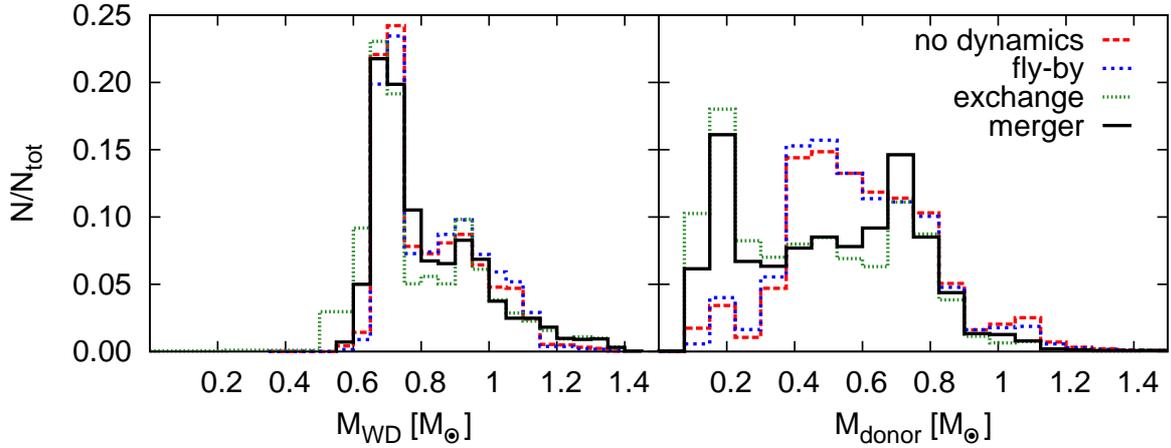}
  \end{center}
  \caption{WD mass (left-hand panel) and donor mass (right-hand panel) distributions 
associated with CVs at their formation times (onset of mass transfer), 
separated according to each formation channel and normalized by the total number
of counts in each formation channel.}
  \label{Fig2}
\end{figure*}

We first quantify the fraction of
primordial CV progenitors that are destroyed in dynamical interactions
before becoming CVs ($f_{\rm dest}$). This is illustrated in 
Fig. \ref{Fig1}, where we show $f_{\rm dest}$ vs. $\Gamma$ for all models,
separated according to the IBP and the CEP efficiency. 
Note that the greater the $\Gamma$, the higher the fraction of destroyed primordial CV progenitors (i.e the stronger the influence of dynamical interactions on destroying these progenitors).
In terms of the soft-hard boundary, the greater the $\Gamma$, 
the shorter the period that defines the boundary between soft and
hard binaries.

One interesting fact is that this correlation is stronger for models
with the Kroupa IBP. Indeed, we carried out Pearson's rank correlation tests, 
and we found strong correlation with more than 99.9 per cent confidence,
being $r=0.78$ and $0.70$, for the Kroupa and Standard IBPs, respectively.

It is not difficult to understand why this correlation is stronger
for models with the Kroupa IBP, and the reason is intrinsically connected
with the period distribution in both IBPs. The majority of the binaries
in the Kroupa IBP have periods longer than $10^3$ days ($\approx$ 83 per cent),
which is not the case for the Standard IBP ($\approx$ 46 per cent)\footnote{In
order to visualise how the period distribution in both IBPs looks like,
readers are recommended to check the appendix in \citet{Belloni_2017b}.}. 
In this way, primordial binaries in the Kroupa IBP are more sensitive with
respect to the strength of dynamics, being much easier affected by interactions
as a whole, since it is predominantly composed of soft binaries, which
is the opposite in the case of the Standard IBP.

After this general overview about CV formation and destruction,
we can turn to the analysis of CV population properties, separated 
according to different formation channels in our models.
In all our models, we identify four main formation channels, namely:
(i) CEP without any influence of dynamics (no dynamics);
(ii) CEP with weak influence of dynamics (fly-by);
(iii) exchange; and
(iv) merger.
More details about how we separate the CVs into these four formation channel groups
are given in \citet{Belloni_2016a,Belloni_2017a,Belloni_2017b}.


\section{CV properties}

\subsection{Progenitor population}
\label{ppCV}

We start the presentation of CV properties by focusing our attention
on the main distributions (i.e. masses, period, eccentricity and 
mass ratio) of CV progenitors.

We note that strong dynamical interactions are able to trigger
CV formation in binaries that otherwise would never undergo a CV
phase, in very good agreement with our previous findings \citep{Belloni_2017a,Belloni_2017b}.
In addition, if we define roughly the range in the parameter space in which primordial CV progenitors 
belong as $M_1/{\rm M}_\odot \in [2,5]$, $q \in (0,0.5]$ and $\log_{10} (P/{\rm d}) \in [2,5]$, we can
compute the fraction of dynamically formed CVs coming from this region.
We find that only $\sim$ 23 per cent of all dynamically formed CV progenitors belong to this
narrow range, which allows us to conclude that, as previously, dynamics extend the parameter space 
applicable to CV progenitors (with respect to CVs formed without influence of dynamics), and allow 
binaries that would not become CVs to evolve into CVs.

\subsection{CVs at the onset of mass transfer}

If dynamically formed CVs have different properties
from CVs formed from primordial binaries, then some of their properties (e.g. component masses and periods) 
at the onset of mass transfer should be different.
Here we discuss two important distributions, the WD mass
and the donor mass.
The WD mass distribution is important because it can
help us to understand the nature of GC CVs, and also because
it has been claimed for a long time that GC CVs have, on
average, more massive WDs when compared with Milky Way (MW) CVs.
The donor mass distribution is important because it
determines the entire CV evolution, including the
mass transfer rate, and in turn the luminosity, 
and the duty cycle.
In this way, CVs with higher donor masses are more likely
to be detected in observational surveys, because they
are bright enough or because they exhibit dwarf nova (DN) outbursts more
frequently.

In Fig. \ref{Fig2}, the WD and donor masses of the CVs at 
the onset of mass transfer are displayed, separated
according to the formation channel.
Note that there is no statistical evidence suggesting that
the WD mass distribution of dynamically formed CVs 
is different from that of CVs formed
from primordial binaries, since the
histograms practically overlap each other.
We stress that this result is in disagreement with our previous works
\citep{Belloni_2016a,Belloni_2017a,Belloni_2017b}, where
we do show that these two sets of CVs are different.

The reason for this discrepancy is associated with
better prescriptions for CV evolution adopted here, 
in particular angular momentum loss prescriptions and 
criteria for dynamically and thermally stable mass transfer,
which makes CVs formed from CEP in our simulations 
to have naturally more massive WDs.

Indeed, we have adopted in all simulations, the eCAML model proposed
by \citet{Schreiber_2016}. According to these authors, if the strength of 
CAML is inversely proportional to the WD mass, then most (if not all) CVs 
with low-mass WDs are dynamically unstable.
The eCAML model is currently the only model that can solve some long-standing problems, like the 
associated CV space density, the period distribution and the WD mass distribution
\citep{Schreiber_2016,Belloni_2018b}. In addition, it can also explain the 
existence of single He-core WDs \citep{Zorotovic_2017}.
The main mechanism thought to be responsible for the postulated dependence of
CAML on WD mass are nova eruptions \citep{Schreiber_2016,Nelemans_2016}.
Such eruptions might cause strong AML by friction that makes most CVs with low-mass WDs 
dynamically unstable, which leads to merger instead of stable mass transfer.
This is because frictional AML produced by novae depends strongly 
on the expansion velocity of the ejecta \citep{Schenker_1998}, and 
for low-mass WDs, the expansion velocity is small \citep{Yaron_2005}.

With respect to the donor mass distribution, we do notice differences
between dynamical and non-dynamical CVs. 
There are relatively more dynamical CVs (formed because of mergers or exchanges)
with donors lighter than $\sim 0.3$ M$_\odot$ 
(formed during interactions with low-mass binaries).
This indicates that dynamics are likely to produce relatively more optically
faint CVs than CEP, at the onset of mass transfer.
%

\begin{figure*}
   \begin{center}
    \includegraphics[width=0.48\linewidth]{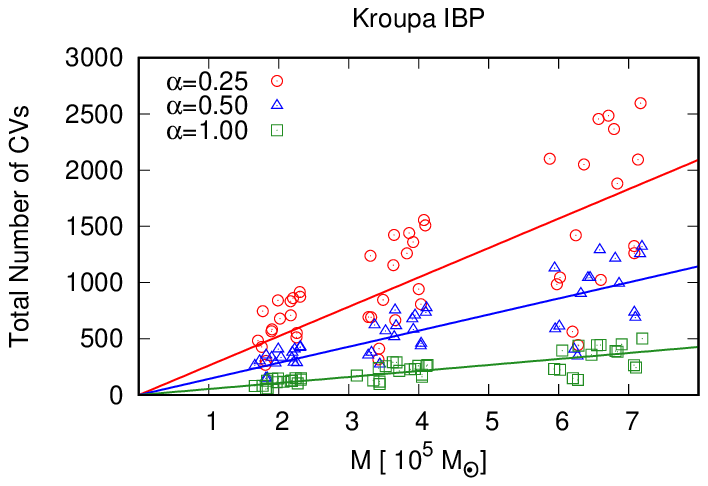}
\hspace{0.2cm}
    \includegraphics[width=0.48\linewidth]{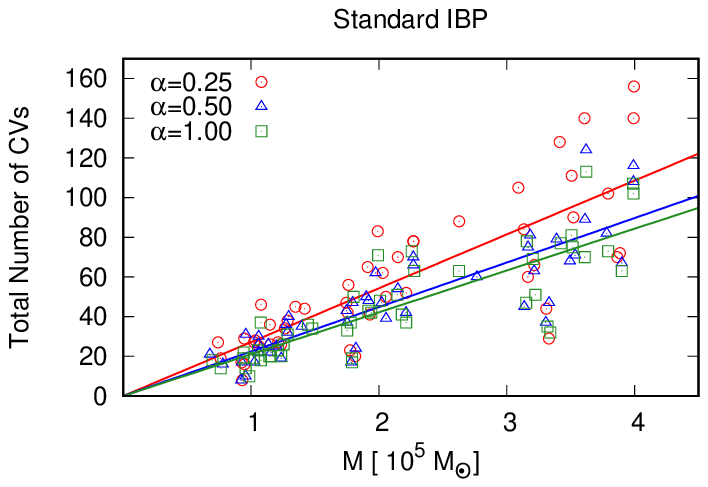}
    \end{center}
  \caption{Total number of present-day CVs in all our models against the cluster mass, 
separated according to the IBP and CEP efficiency. The lines are linear regression 
models ($N_{\rm tot} = a \times M$, where $N_{\rm tot}$ and $M$ are the 
total number of present-day CVs and the present-day cluster mass in $10^5$~M$_\odot$, respectively). 
Note that the value of the CEP efficiency significantly changes the number of CVs in 
models following the Kroupa IBP. Values of $a$ are provided in Table \ref{TabCoef}.}
  \label{Fig10}
\end{figure*}


\subsection{Present-day CV population}
\label{pdp}

One interesting correlation one would expect is related
to the number of CVs and GC masses, i.e. the greater the 
GC mass, the greater the number of CVs.
We found here that this correlation holds for all combinations
of IBP and CEP efficiency, as illustrated in Fig. \ref{Fig10}.

The total amount of present-day CVs in all 288 models is 96214,
being $\approx 92.5$ per cent coming from models set with the Kroupa IBP and
only $\approx 7.5$ per cent from those set with the Standard IBP.
Regarding the CEP efficiency, as expected, the smaller the $\alpha$,
the greater the number of CVs, i.e. models set with $\alpha=0.25$, $0.5$ and
$1.0$ contribute with $\approx 56.5$, $\approx 30.5$ and $\approx 13.0$ per cent, 
respectively.

\begin{table}
\centering
\caption{Coefficient for the best-fitting lines in the plane total number of CVs
versus present-day cluster mass, in the form of 
$N_{\rm tot}~=~a~\left(M/10^5{\rm M}_\odot\right)$, for
all models in our simulations grouped according to the IBP (Kroupa or Standard)
and the CEP efficiency ($\alpha=$0.25, 0.50 and 1.00). The lines are
plotted in Fig. \ref{Fig10}.}
\label{TabCoef}
\begin{adjustbox}{width=0.99\linewidth}
\noindent
\begin{threeparttable}
\noindent
\begin{tabular}{lcccccc}
\hline\hline
 & & \multicolumn{5}{c}{Kroupa IBP}   \\
\hline
$\alpha$ & & 0.25 & & 0.50 & & 1.00 \\ 
\hline
$a$  &  & $266.1 \pm 14.9$ & & $143.0 \pm 6.3$  & & $53.4 \pm 2.3$   \\
\hline
\hline
 & & \multicolumn{5}{c}{Standard IBP}   \\
\hline
$\alpha$ & & 0.25 & & 0.50 & & 1.00 \\ 
\hline
$a$  &  & $27.1 \pm 1.3$ & & $22.4 \pm 0.9$  & & $21.1 \pm 0.9$   \\
\hline\hline
\end {tabular}
\end{threeparttable}
\end{adjustbox}
\end{table}

In order to provide readers a way to estimate the amount of CVs
per GC mass, we performed linear regressions for the number of CVs
against the host GC mass, in the form 
$N_{\rm tot}~=~a~\left(M/10^5{\rm M}_\odot\right)$,
for all models in our simulations grouped according to the IBP
and the CEP efficiency. 
The best-fitting coefficients are provided in Table \ref{TabCoef}
and the best-fitting lines are plotted in Fig. \ref{Fig10}.

From Fig. \ref{Fig10}, we clearly see that the number of CVs significantly
changes with regards to $\alpha$, for models set with the Kroupa IBP. 
On the other hand, for models set with the Standard IBP, the number of CVs 
is almost insensitive with respect to $\alpha$. 
This result is a direct consequence of the respective period distributions.
As the period distribution in the Kroupa IBP smoothly increases towards longer
periods, the smaller the alpha, the higher the amount of CVs that survive
the CEP, since lower values for $\alpha$ lead to shorter periods after CEP.
In this way, since the number of binaries increases with period in the Kroupa IBP,
more and more binaries manage to become CVs due to the fact that the pre-CV 
lifetimes are reduced, as the value of $\alpha$ becomes smaller.
Now, for the Standard IBP, since the period distribution is log-uniform throughout
the entire range of period, the effect of moving the range from which CV progenitors
belong towards longer periods (by decreasing $\alpha$) has only a small effect on
the amount of CVs formed.

In what follows, we concentrate only on observational properties
related to GC CVs.
In other words, hereafter we only investigate properties of
present-day CVs that are likely to be observed via
multiple technique methods, since these are the most important 
ones and can potentially lead to some constraints.

\begin{figure*}
   \begin{center}
   	\includegraphics[width=0.48\linewidth]{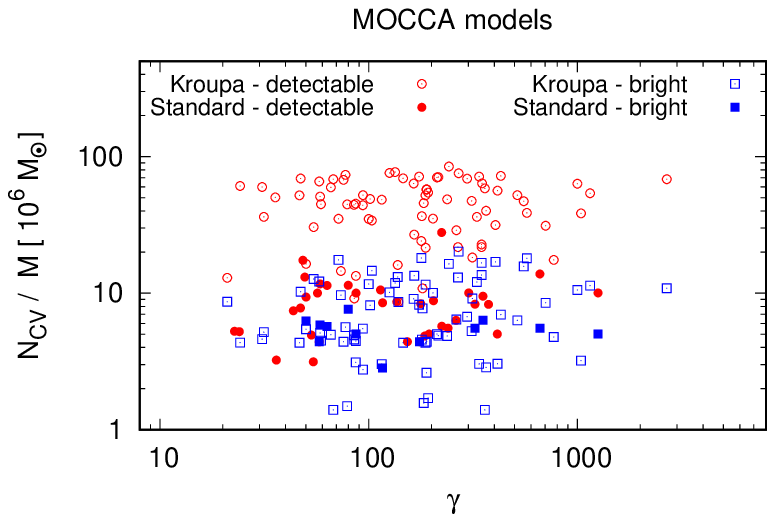}
    \includegraphics[width=0.48\linewidth]{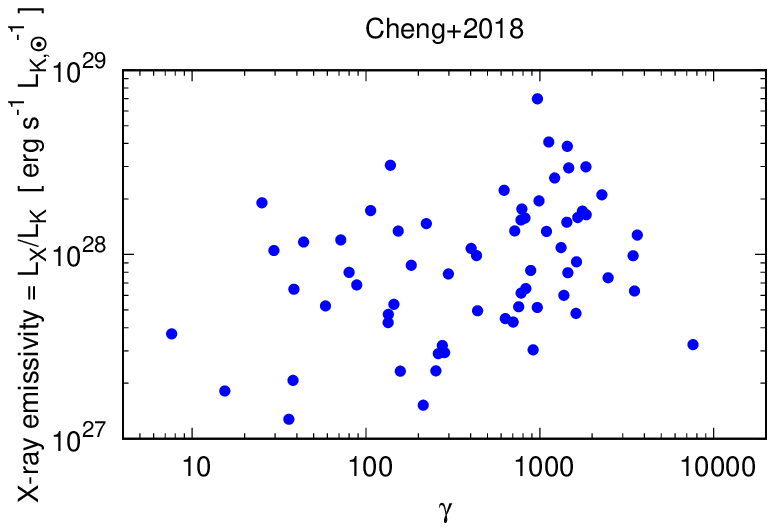}
    \end{center}
  \caption{In the left-hand panel, we show results of our models for the number of both, detectable and bright CVs per unit mass
vs. the mass-normalized stellar encounter rate ($\gamma$). We separated the models with respect to the IBP (i.e. Kroupa and Standard).
In the right-hand panel, we show observational results by \citet{Cheng_2018}.
The X-ray emissivity,
defined as the ratio between the X-ray luminosity and the K-band 
luminosity $(L_X / L_K)$, is considered a measure of the number of X-ray source abundance in a GC (in the study of these authors, mainly CVs and ABs).
Note that, in both panels, there is no (if at all very weak) statistical 
evidence for a correlation between the CV abundance or faint ($L_X \lesssim 10^{34}$ erg s$^{-1}$) X-ray sources and $\gamma$.}
  \label{FigGamma}
\end{figure*}


\section{Present-day CV Population: detectable CVs}
\label{detec}

\subsection{Criteria to be considered detectable}

So far we know that core-collapsed GCs show 2 populations of CVs \citep{Cohn_2010,Lugger_2017}, and that based on that we can separate the CV population for non-core-collapsed GCs \citep{Cool_2013,Rivera_2017}.
In order to investigate whether we can reproduce
some observational features of these two populations, we first
select the detectable CVs in our models\footnote{We stress that we removed
from the population of detectable CVs unstable and extremely young systems. The former is because these CVs will quickly merge and likely do not
contribute to the observed population. The latter is
because the mass transfer rate provided by the \bse code is not reliable
when the CV has just born.}. 
To do so, we assume a conservative definition in order
to allow for statistical analysis and to be consistent
with observations. We used as cut-offs the
donor mass, X-ray luminosity and the absolute visual magnitude,
being the last two computed as described in \citet{Belloni_2016a}
and assuming that the average accretion rate onto the WD during quiescence
is 1 per cent of the average mass transfer rate\footnote{We emphasize that
the accretion rate onto the WD during quiescence is not constant
and depends on the CV properties (e.g. WD mass, mass transfer rate, etc). 
Our choice of having the average accretion rate being 1 per cent of the 
average mass transfer rate is then an approximate estimate, 
but consistent with results from the disc instability model \citep[e.g.][]{Lasota_2001}.}.
Our limiting quantities are: donor mass is $0.1$ M$_\odot$,
X-ray luminosity is $5\times10^{29}$ erg s$^{-1}$
and the limiting absolute visual magnitude is 14 mag 
(i.e. 10 mag below the turn-off magnitude). 
These three values are consistent with observational limits \citep{Cohn_2010,Cool_2013,Lugger_2017,Rivera_2017,Henleywillis_2018}.

From the total of 96214 present-day CVs in all our simulations,
after applying these criteria, we have 2129 detectable CVs.
This provides that, on average, only between $2-4$ per cent (depending on the CEP efficiency) 
of the CVs in a GC can eventually be detected.

We now separate the detectable CVs between bright and
faint.
Bright CVs presumably have their optical fluxes dominated by the donor. 
Whereas faint CVs have theirs dominated by the WD and/or accretion disc. 
To be consistent with observations, we adopt here a cut-off based on the 
absolute visual magnitude, which is defined by $M_V$=9 mag.
Detectable CVs which absolute visual magnitudes
are smaller than that are bright CVs, and they are faint otherwise.
We emphasize that this criterion is suitable for our purposes,
since our goal is to infer statistical properties from our
models rather than modelling particular GCs. 

After filtering out the CVs with respect to limiting luminosities
and separating them according to their brightness, an additional
and final criterion has to be applied, regarding the position
in the cluster. In observations, the observed region is usually
within the half-light radius ($R_L$), and for this reason, in
some comparisons with observations, we also separate 
them with respect to $R_L$.

\subsection{Influence of the cluster type}

With respect to the cluster type, for those models set with the 
Kroupa IBP, we notice that most core-collapsed models have
fractions of bright CVs with respect to detectable CVs in the range 
of $5-45$ per cent, and most non-core-collapsed models have 
them in the range of $7-33$ per cent.
However, a small portion of our core-collapsed models have more 
than $50$ per cent of bright CVs.
These models are compact and characterized by high values of the
central surface brightness, and short half-mass relaxation times. 
These models have then properties much closer to real core-collapsed 
GCs than the whole set of our core-collapsed clusters.

On the other hand, for models set with the Standard IBP,
most of them have none bright CVs ($\sim90$ per cent of them),
and only a few have non-null fractions of bright CVs.
We can conclude then that, in general,
such an IBP cannot reproduce observed fractions of bright 
CVs among core-collapsed and non-core-collapsed clusters.

The fractions of bright CVs we found for models set with 
the Kroupa IBP are, in general, consistent 
with observations of non-core-collapsed GCs ($\sim20-25$ per cent).
They are also consistent with respect to core-collapsed GCs 
($\sim40-60$ per cent), provided we take into account only 
models whose properties are much closer to observed core-collapsed GCs.

Regarding the detectable CV spatial locations, we found that,
on average, only $\sim 45$ per cent  
of detectable CVs (both bright and faint) are inside the half-light radius, 
which corresponds to $\sim 42$ and $\sim 47$ per cent 
of bright and faint CVs, respectively, inside it.
However, such fractions should be considered upper limits as
they depend on the cluster properties. In particular,
as shown in Section \ref{spadist}, such fractions of detectable CVs 
inside/outside $R_L$ strongly depends on the cluster half-mass relaxation times.

\subsection{Are most CVs dynamically formed?}

With respect to the formation channels, the dominant one amongst detectable CVs 
is typical CEP  ($\approx88^{+12}_{-18}$ per cent, for both core-collapsed and 
non-core-collapsed clusters).
We also found that the average fraction of dynamically formed CVs among only bright
CVs is relatively low ($\approx9_{-9}^{+24}$ per cent, for both core-collapsed and 
non-core-collapsed clusters).
In other words, we found here no (if at all very weak) correlation between the 
number of either detectable CVs or bright CVs with respect to the cluster type
(e.g. related to the stellar encounter rate).

Our results are in disagreement with 
previous conclusions that bright CVs 
were predominantly dynamically formed (via exchange), 
and faint CVs were a mix of CVs formed in different channels 
\citep[][]{Belloni_2017b,Hong_2017}.
This is likely connected with the high (non-realistic) CEP 
efficiency adopted previously, since the higher the CEP efficiency,
the smaller the number of CVs formed from primordial binaries,
especially for models following the Kroupa IBP.
In this way, the contribution from primordial CVs have been
underestimated in our previous works.

We note that our current findings are in agreement with recent studies of 
\textit{Chandra} X-ray sources in GCs by \citet{Cheng_2018}. 
Using a sample of 69 GCs and focussing on CVs and chromospherically
active binaries (ABs),
these authors found that there is not a significant correlation between the number of X-ray sources 
and the mass-normalized stellar encounter rate in units of $10^6~\Msun$ ($\gamma=\Gamma/M_6$). 
These findings disagree with previous results, which considered smaller GC samples \citep[e.g.][]{Pooley_2006}. 
A correlation would be expected if dynamical interactions largely influence the creation of X-ray sources. However, 
\citet{Cheng_2018} have shown that dynamical interactions are less dominant than previously believed, and that the primordial formation has a substantial contribution. 
In Fig. \ref{FigGamma} we show the number of CVs as a function of $\gamma$. 
We show numbers for both, detectable
and bright CVs normalized by the total cluster mass in units of $10^6~\Msun$. 
In that figure we also show results from \citet{Cheng_2018} for the X-ray emissivity 
(defined by these authors as the X-ray luminosity divided by the K-band luminosity and
considered a measure of the X-ray source abundance, mainly CVs and ABs)  vs. $\gamma$. 
Note that both observational and theoretical results show
no (if at all very weak) statistical evidence for a correlation 
between the normalized CV abundance and $\gamma$.

The physical reason for that is associated with the role of dynamics
in creating and destroying pre-CVs.
We notice that destruction of CV progenitors take part mainly for 
MS-MS binaries during the first few hundred Myr of cluster evolution. 
Later, when MS-WD binary is created, dynamical interactions are very 
strongly suppressed, because during the CEP there is a substantial 
reduction of binary periods.

Regarding dynamical pre-CV formation, one can have mainly three possible 
scenarios:

i) {\it interaction between a low mass MS-MS binary and single C/O WD}:
this type of interaction would lead to exchange interaction in which
a MS is replaced with a WD.
Such an interaction has to form rather compact binary, since the binary
evolution needs about $\lesssim10$ Gyr to make it a CV. If the exchanged 
binary is wider (say two times), then it will need several dynamical 
interactions to bring it to the period which will result in CV formation. 
Since the Spitzer's average change of binding binary energy is 20 per cent
\citep{Spitzer_BOOK}, it would be needed around four such interactions, 
which is rather not probable for large number of such binaries.

ii) {\it interaction between a low mass MS-MS binary and single MS}:
this interaction would lead to an MS-MS binary which, before WD formation, 
is brought to the adequate range (Section \ref{ppCV}) that will guarantee 
CEP and further binary evolution (during $\sim10$ Gyr) leading to CV formation.
We found that the average number of such interactions is only
$\sim0.05$ per CV, which clearly shows that this channel is not important.

iii) {\it interaction between a WD-MS binary and single MS}:
there are two posibilities here, either
this interaction is strong and generates a pre-CV via the 
replacement of the MS in the binary with a more massive MS intruder,
or a few interactions occur which harden such a WD-MS binary to become
a pre-CV.
For this scenario, the probability is also low, provided the
small average semi-major axis of WD-MS binaries ($\sim10 \Rsun$)
and typical properties of GCs inferred from the models.
Indeed, the average numbers of weak and strong interactions associated
with pre-CV binaries over the time-scale of $\sim10$ Gyr (time
between WD-MS binary and CV formations) are $\sim0.126$ and $\sim0.003$ 
per CV, respectively. 

To sum up, we show that the main scenarios proposed in the literature 
\citep{Ivanova_2006,Shara_2006,Belloni_2016a,Belloni_2017a,Belloni_2017b,Hong_2017}
for dynamical formation of faint and bright
CVs in GCs have very low probability to happen, which explains our
findings with respect to the influence of dynamics in CV formation
(very low fraction of dynamically formed faint and bright CVs) and 
with respect to the stellar encounter rate (no/extremely weak correlation 
with the amount of detectable and bright CVs).
We notice that our explanation is supported by recent observations \citep{Cheng_2018},
which show that there is no (or very weak) correlation between faint ($L_X \lesssim 10^{34}$ erg s$^{-1}$) X-ray 
sources, presumably mainly composed of CVs and ABs, and $\gamma$.

\begin{figure*}
   \begin{center}
    \includegraphics[width=0.95\linewidth]{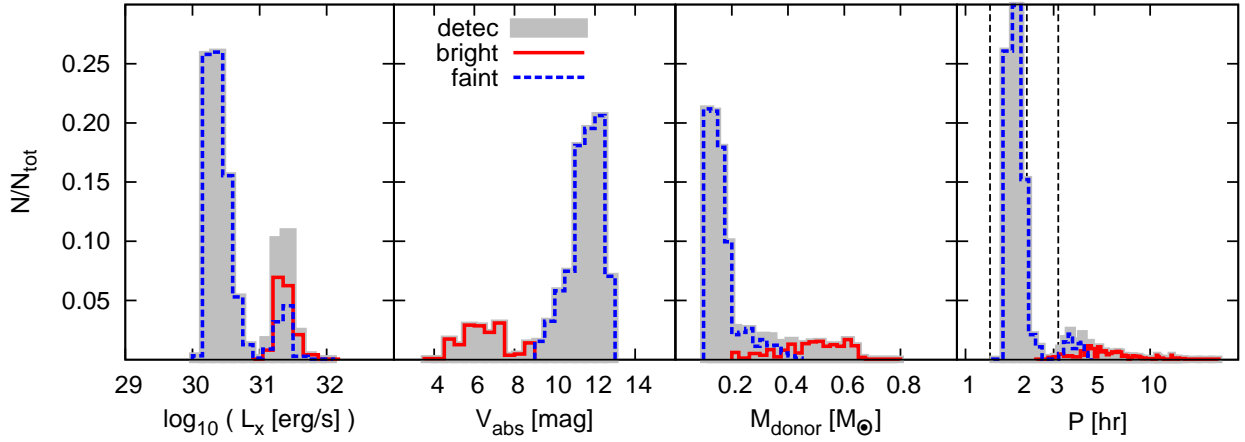}
    \end{center}
  \caption{X-ray luminosity, absolute visual magnitude, donor mass and
period distributions associated with detectable CVs (i.e. bright and faint CVs).
Each histogram was normalized by the total amount of detectable CVs.
Vertical lines in the period distribution are the observational 
location of the period minimum \citep{Gansicke_2009} 
and gap edges \citep{Knigge_2006}. Notice that bright CVs are mainly above
the gap (with some inside it) and most faint CVs are below
the gap (with a few right above it).
}
  \label{Fig3}
\end{figure*}

\begin{figure*}
   \begin{center}
    \includegraphics[width=0.45\linewidth]{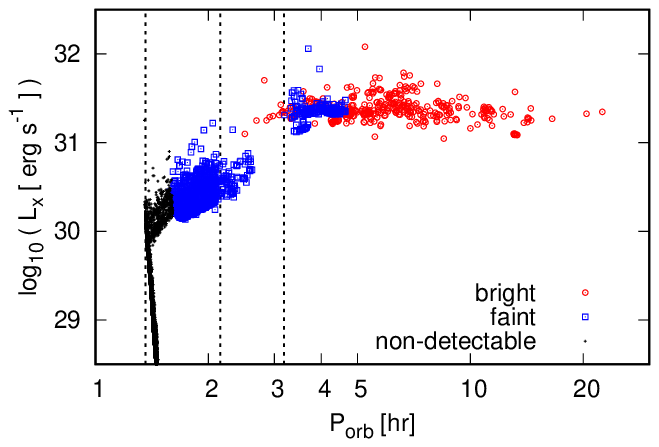}
    \includegraphics[width=0.45\linewidth]{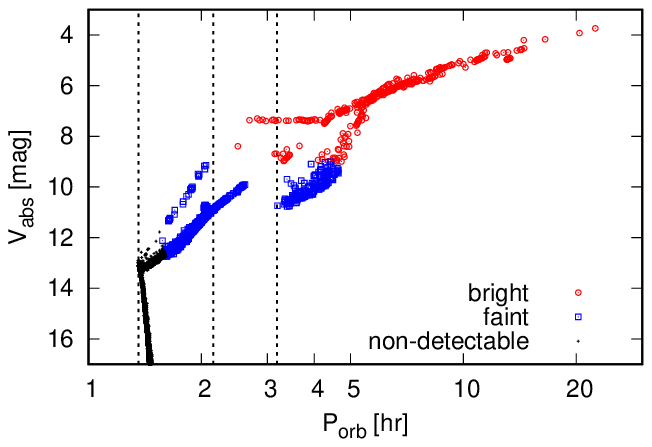}
    \includegraphics[width=0.45\linewidth]{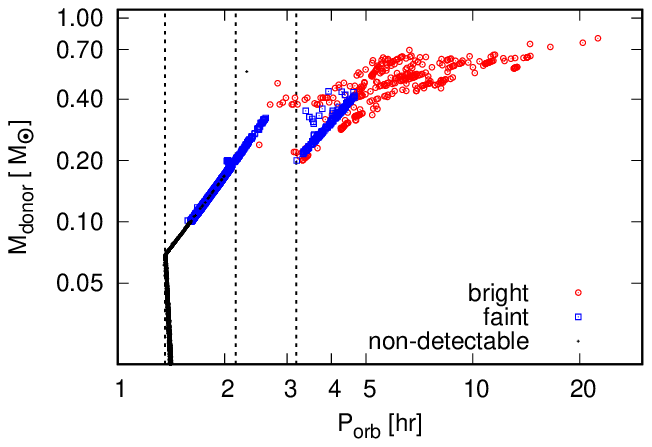}
    \includegraphics[width=0.45\linewidth]{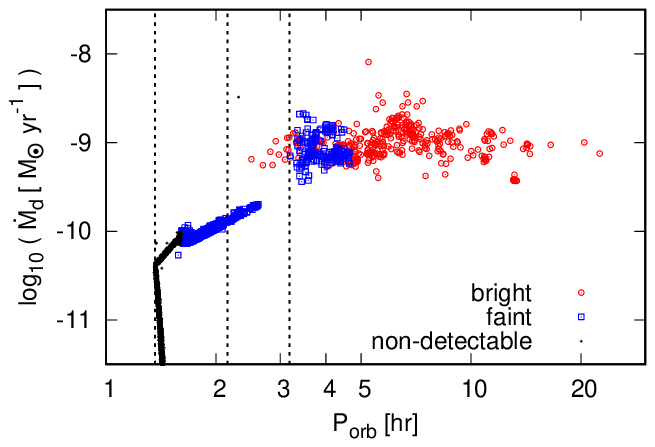}
    \end{center}
  \caption{Some CV properties against period: X-ray luminosity (top left-hand panel), 
absolute visual magnitude (top right-hand panel),  donor mass (bottom left-hand panel) 
and mass transfer rate (bottom right-hand panel). Vertical lines are the observational 
location of the period minimum \citep{Gansicke_2009} and gap edges \citep{Knigge_2006}.}
  \label{Fig4}
\end{figure*}

\begin{figure*}
   \begin{center}
    \includegraphics[width=0.95\linewidth]{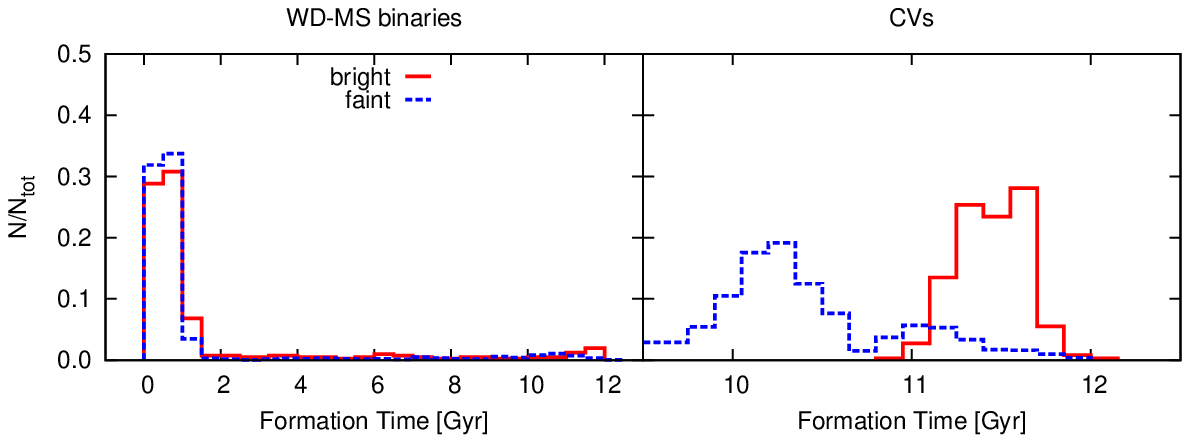}
    \includegraphics[width=0.95\linewidth]{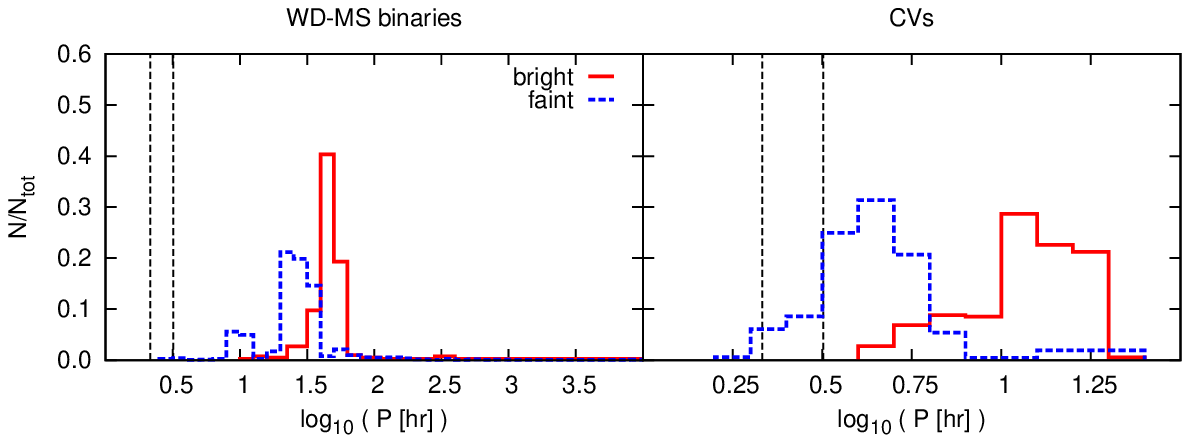}
    \end{center}
  \caption{Normalized formation rates (top row) of WD-MS binaries that will later become the 
detectable CVs (left-hand panel) and of detectable CVs (right-hand panel), and their
period distributions (bottom row).
Vertical lines in the period distributions are the observational 
location of the  gap edges \citep{Knigge_2006}.
Note that most WD-MS binaries (pre-CVs) are formed before $\sim~1$~Gyr and most
CVs are formed after $\sim~10$~Gyr.}
  \label{Fig5}
\end{figure*}

\subsection{Orbital, photometric and X-ray properties}

In order to check whether we are able to reproduce the bimodality of GC CVs, 
we show in Fig. \ref{Fig3} the distributions of the absolute visual magnitude,
X-ray luminosity, donor mass and period for all detectable CVs.
Note that we have indeed evidences towards a bimodal population, even though
the population of faint CVs is clearly dominant in the four distributions.
In particular, it is interesting that we find a bimodality
in the X-ray luminosity distribution, where we have a population of faint
X-ray CVs ($L_X \lesssim 10^{31}$ erg/s) and another population
of bright X-ray CVs ($L_X \gtrsim 10^{31}$ erg/s), which actually is in good agreement with observations \citep[][see their fig. 14]{Rivera_2017}. However, we note that detectable CVs with X-ray luminosities below $10^{30}$ erg/s are missing in our distribution, whereas CV candidates have been observed below that limit.   
This observed population of very faint X-ray CVs likely have extremely low-mass donors ($<0.1$~M$_\odot$), and periods close to the period minimum 
(period bouncers or their progenitors). In fact, some Galactic field CVs, such as GW Lib and WZ Sge, which have X-ray luminosities $0.05^{+0.10}_{-0.02}\times10^{30}$ and 
$0.7^{+0.3}_{-0.1}\times10^{30}$ erg s$^{-1}$ [2-10 keV], respectively 
\citep{Byckling_2010}, seem to support this explanation.
Alternatively, we notice that such X-ray luminosities could also be reached 
in cases where the time-averaged accretion rate in short-period CVs would be 
smaller than the one assumed here 
(i.e. $<~10^{-2}~\langle\dot{M}_{\rm tr}\rangle$, where 
$\langle\dot{M}_{\rm tr}\rangle$ is the time-averaged mass transfer rate).

Regarding the period distribution of the detectable CVs, we see that bright systems dominate the distribution above the period gap. Whereas the faint systems have periods shorter than 4h, with a vast majority below the period gap. Note, however, that the gap is populated by both types of CVs. 
In Fig. \ref{Fig4}. we show how the orbital period of the detectable and non-detectable
CVs are related to the  X-ray luminosity, absolute visual magnitude,  
donor mass and mass transfer rate. From this figure we see that bright CVs have longer periods, higher mass transfer rates, higher
donor masses and higher luminosities. This is in general agreement with the properties found for the bright CVs in the 4 GCs that we are considering, which have periods longer than 2.4 h \citep[see e.g.][and references therein]{1996_bailyn,2003_kaluzny,Rivera_2017}. Unfortunately CVs below the period gap have not been confirmed in any of these clusters. 
From Fig. \ref{Fig4} we can also clearly see the typical trend associated with CV evolution, which goes
from long to short periods. Note that due to our chosen mass limit for detectable CVs (M$_{\rm donor}>0.1$~M$_\odot$), we miss in these plots the portion of extremely faint
CVs composed of period bouncers. Our results then suggest that those CV candidates in NGC 6397, NGC 6752, 47 Tuc and $\omega$ Cen
with very low luminosities, are likely period bouncers or CVs very close to the period minimum (WZ Sge-type).


\begin{figure*}
   \begin{center}
    \includegraphics[width=0.95\linewidth]{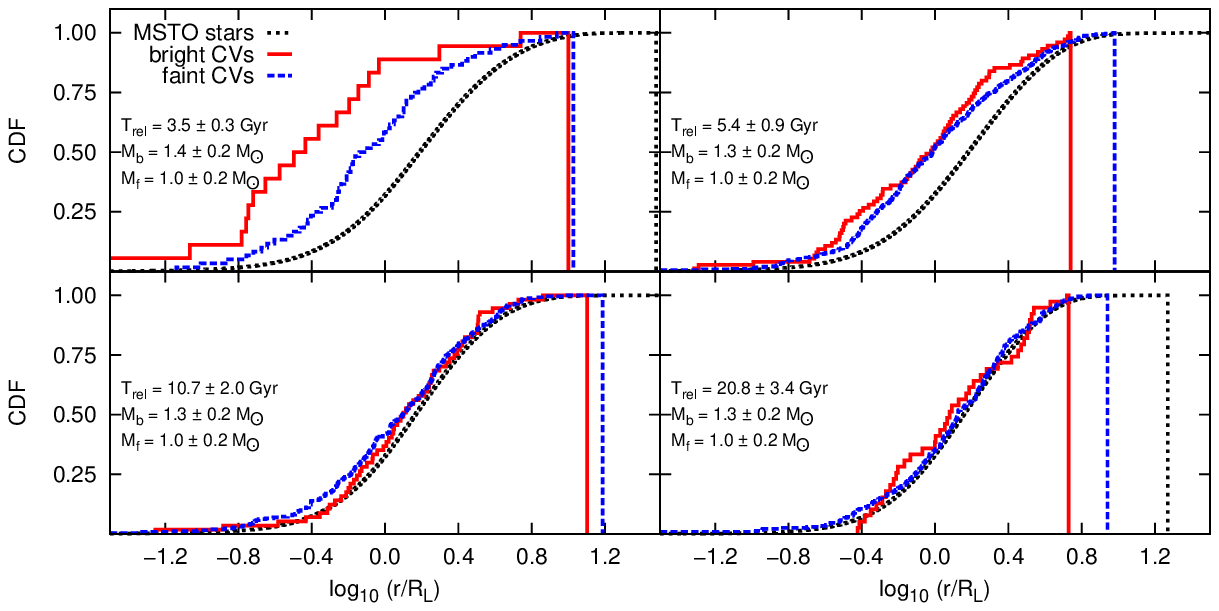}
    \end{center}
  \caption{Cumulative radial distribution function (CDF) for detectable CVs
(faint and bright), for clusters grouped according to different half-mass relaxation
times ($T_{\rm rel}$):  
 $T_{\rm rel}/$Gyr < 4 (top left-hand panel),
4< $T_{\rm rel}/$Gyr < 8 (top right-hand panel),
8< $T_{\rm rel}/$Gyr < 15 (bottom left-hand panel),
$T_{\rm rel}/$Gyr > 15 (bottom right-hand panel).
We also indicate in each panel the mean values of $T_{\rm rel}$, and
bright and faint CV masses ($M_{\rm b}$ and $M_{\rm f}$, respectively).
Note that, mean values of $M_{\rm b}$ and $M_{\rm f}$ are similar in all panels and
that, for different $T_{\rm rel}$, bright and faint 
CVs show different levels of concentration relative to the MSTO stars.}
  \label{Fig6}
\end{figure*}

\begin{figure}
   \begin{center}
    \includegraphics[width=0.98\linewidth]{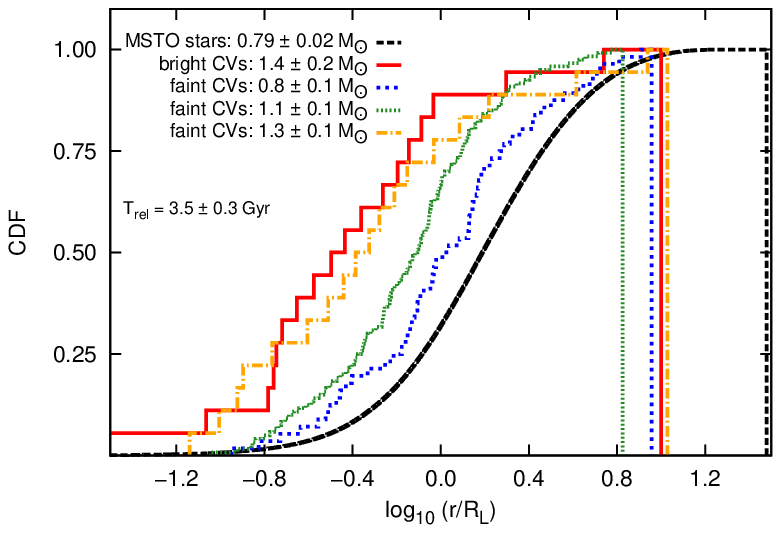}
    \end{center}
  \caption{Cumulative radial distribution function (CDF) for detectable CVs
(faint and bright), for clusters which half-mass relaxation times ($T_{\rm rel}$) 
are shorter than 4 Gyr.
We separated the faint CVs into three groups, according to their masses,
namely (i) $M<0.8$ M$_\odot$ (blue curve), (ii) $0.8<M/$M$_\odot<1.2$ (green curve),
and (iii) $M>1.2$ M$_\odot$ (orange curve).
In the keys we provide the average mass of each stellar group, i.e. MSTO stars,
bright CVs, and faint CVs separated as described above.
Note that, different faint CV masses show different levels of concentration
relative to the MSTO stars.}
  \label{Fig7}
\end{figure}

\subsection{Spatial Distribution: general features}
\label{spadist}

From the discussion so far, it is clear that the
dominant formation channel amongst bright and faint CVs is the typical
CEP, as in MW CVs. In this way, a natural follow-up concern is how to
explain their properties, including their spatial distributions\footnote{We 
note that some insight has been given by \citet{Hong_2017} about the spatial
distribution of CVs in GCs but here we provide a more detailed explanation.}.
Interestingly, \citet{Cohn_2010} and \citet{Lugger_2017}
suggest that their findings are consistent with an evolutionary 
scenario in which CVs are produced by dynamical interactions near the cluster 
centre and diffuse to larger orbits as they age.
However, the results of our simulations suggest that this scenario 
is not very likely.
Indeed, a CV that is dynamically formed in a cluster core is ejected outside the central parts due to the strength
of the dynamical interaction. Thus, it is unlikely that it will remain in the core after formation.
In addition, the probability for a CV to interact is
extremely small, given their short periods \citep{Leigh_2016,Belloni_2017a}. 
In this way, scattering interactions acting to
increase their orbits and thus, forcing their migration to larger radii in the 
cluster potential are not very likely.
Finally, such a scenario seems to ignore completely the contribution
of primordial CVs to the present-day CV population.

In order to provide readers a scenario which is compatible with our findings, 
we should consider the WD-MS binary and CV formation times, their properties at 
the formation times, and the cluster $T_{\rm rel}$.

In Fig. \ref{Fig5}, we show the normalized formation rate of CVs and WD-MS binaries
that are their progenitors, separating them into bright and faint populations. 
Note that most WD-MS binaries are formed before $\sim 1$ Gyr, which is the time
for the formation of most C/O WDs, given the metallicity adopted here (Z=0.001). 
The remaining WD-MS binaries that are formed after this time are mainly dynamical ones.
On the other hand, the detectable CVs are formed mainly after $\sim 10$ Gyr,
which is more or less the limiting time for them to be bright enough to be
detected at the present day (i.e. at 12 Gyr).
In this way, the time required for a CV to evolve beyond the detection limit
is around $\sim 2$ Gyr, which is consistent with the time-scale associated
with long-period CVs to evolve towards close to the period minimum (donor
mass of $\lesssim 0.1$ M$_\odot$).

Said that, we reach the following general picture related to the origin
of faint and bright CVs.
Since most of them are formed via CEP under no/weak influence of dynamics and
since CVs have C/O WDs\footnote{We note that for a long time, many low-mass He WDs have been predicted to be found in CVs, but not a single system with a definite He-core WD
has been identified so far \citep{Zorotovic_2011}. This inconsistency between modelling and
observations can be overcome with the eCAML model proposed by \citet{Schreiber_2016}, 
which is adopted here.}, 
the WD-MS binaries should have to be formed during the early cluster evolution,
but they should have periods such that systemic angular momentum losses should
take more than $9$ Gyr to shrink their orbits such that they become interacting
binaries, i.e. CVs. This is precisely the case as seen in Fig. \ref{Fig5}.
Note that most WD-MS binaries that are progenitors of bright CVs 
are born with periods longer than $\sim 30$ hr. On the other hand,
most of those that are faint CV progenitors are born with periods 
between $\sim 10$ and $\sim 30$ hr. With such long periods, it is
not surprising that they take more than 9 Gyr to become CVs.

In this picture, the mass segregation comes naturally, since the WD-MS binaries
are very hard to interact strongly during the cluster evolution and have their 
properties changed \citep[e.g.][]{Leigh_2016,Belloni_2017a}, 
but they could have enough time to sink towards the core, if the $T_{\rm rel}$ is short enough to allow that.
This occurs for both bright and faint CVs, as their masses (including both components, the WD and MS star) 
are, on average, $\sim 1.42\pm0.19$ and $\sim 1.11\pm0.21$ M$_\odot$, respectively, 
i.e. they are more massive than the cluster average mass within the half-mass radius.
We emphasize that this would happen if the cluster half-mass relaxation is substantially shorter than the Hubble time, 
especially because the WD-MS progenitors are initially amongst the most massive objects in the cluster
and segregation starts already at the very beginning.

In general, the shorter the $T_{\rm rel}$, the faster the 
mass segregation within such a cluster. 
In addition, as bright and faint CVs are more massive than MSTO stars, the shorter 
the $T_{\rm rel}$, the more the mass segregation of CVs with respect to
these stars.
If $T_{\rm rel}$ is short enough, say $\lesssim$ 3 Gyr, then differences between 
bright and faint CVs also become clear, since bright CVs are, in general, more massive than 
faint CVs.
Indeed, at the present day, faint CVs are less massive either because they already evolved
as CVs or because their donor masses are initially small and the onset of mass transfer
is close to the present day.
Thus, in the case when the cluster half-mass relaxation time is much shorter than the Hubble time, 
we should expect stronger differences between bright and faint CV spatial distributions,
being bright CVs much more segregated than the faint ones.
%

\begin{figure}
   \begin{center}
    \includegraphics[width=0.95\linewidth]{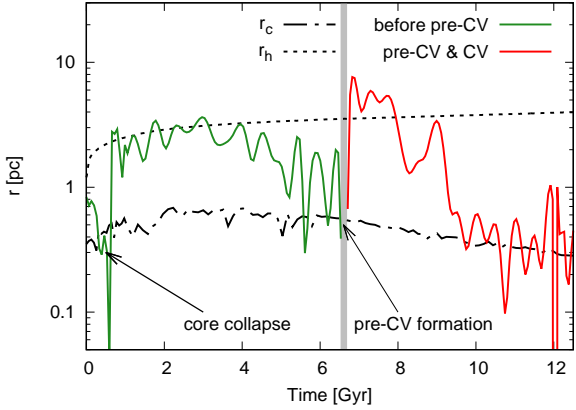}
    \end{center}
  \caption{Illustrative CV radial position evolution, 
  together with the host cluster core and half-mass radii evolution.
  The cluster evolves towards core collapse until $\approx0.5$ Gyr when 
  the core bounce occurs, the pre-CV is
  formed due to dynamical exchange at $\approx6.7$ Gyr and the
  CV is formed close to the present day.
  Note that due to the strong interaction, the newly formed pre-CV
  is expelled from the central parts, but returns to it due to 
  mass segregation.
}
  \label{FigR}
\end{figure}

In order to better understand how CV spatial distribution depends on
the cluster half-mass relaxation time, we first separate all models having initially 1200k 
objects (due to their higher number of CVs) into four groups
according to their relaxation times, namely 
(i) $T_{\rm rel}/$Gyr < 4,
(ii) 4< $T_{\rm rel}/$Gyr < 8,
(iii) 8< $T_{\rm rel}/$Gyr < 15,
(iv) $T_{\rm rel}/$Gyr > 15.
In Fig. \ref{Fig6} we depict the cumulative radial distribution function for 
detectable CVs (faint and bright), for clusters grouped according to the above-mentioned
ranges of $T_{\rm rel}$.
Note first that the average mass of bright and faint CVs, in all panels, are quite
similar, being $\sim 1.4\pm0.2$ and $\sim1.0\pm0.2$ M$_\odot$ so that differences
in mass segregation should come mainly because of differences in $T_{\rm rel}$.
In the bottom right-hand panel, $T_{\rm rel}$ is much longer than the
Hubble time, which implies that there is no visible difference between CVs and
MSTO stars, since they do not have enough time to separate inside the cluster.
In the bottom left-panel, $T_{\rm rel}$ is slightly shorter than the Hubble time
and there is a small difference between CVs and MSTO stars.
As $T_{\rm rel}$ decreases further, the difference between CVs and MSTO stars
becomes more and more clear, as illustrated in top right-hand panel, where
$T_{\rm rel} \sim 5-6$ Gyr.
By decreasing even further, differences between bright and faint CVs become
more and more clear, as seen in top left-hand panel for clusters whose half-mass
relaxation times are $\sim 3-4$ Gyr.
Thus, to sum up, the shorter the $T_{\rm rel}$, the faster and efficient
the mass segregation of bright CVs with respect to faint CVs, and faint
CVs with respect to MSTO stars.

Now, in order to better understand how the mass segregation changes
with respect to the faint CV masses, we show in Fig. \ref{Fig7}
the cumulative radial distribution function of bright and faint CVs for clusters whose 
$T_{\rm rel}~<~4$~Gyr, separating the faint CVs into three groups, according to their masses,
namely 
(i) $M_f<0.8$~M$_\odot$ (blue curve), 
(ii) $0.8<M_f/$M$_\odot<1.2$ (green curve),
and 
(iii) $M_f>1.2$~M$_\odot$ (orange curve).
Note that the smaller the faint CV masses, the weaker the mass
segregation with respect to MSTO stars.
If $T_{\rm rel}$ is conveniently short,
as faint CVs evolve and become less massive, their level of segregation will 
quickly adjust to that of MSTO stars. 
In this way, the `speed' of segregation of faint CVs decreases as they evolve and 
reduce their masses. In any event, they keep segregating, but not as `fast' as before,
because they are less massive now. In the case they have masses similar to MSTO
stars, they should segregate in a similar fashion. In the case
they are more massive, then they would segregate more than MSTO stars.
Thus, not only the cluster $T_{\rm rel}$ is important for mass segregation, but also 
the CV masses and the rates of mass decrease.

\begin{figure*}
   \begin{center}
    \includegraphics[width=0.99\linewidth]{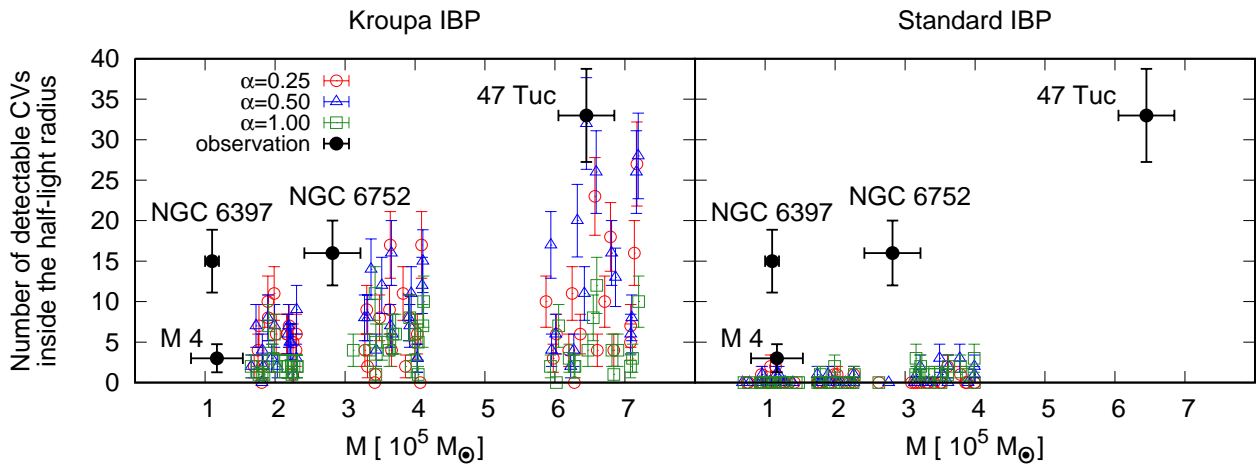}
    \end{center}
  \caption{Number of detectable CVs in our models against the cluster mass, 
separated according the IBP and CEP efficiency. Observational numbers
obtained from \citet[][M 4]{Bassa_2004}, 
\citet[][NGC 6397]{Cohn_2010}, 
\citet[][NGC 6752]{Lugger_2017} and 
\citet[][47 Tuc]{Rivera_2017}. 
For consistency with the other clusters, in the case of 47 Tuc the plotted number 
corresponds to the CVs detected in the optical filters and not to the 
whole NUV CV sample reported by these authors. 
Observational masses were extracted from 
\citet[][]{Kimmig_2015} and \citet{Heyl_2012}.
Vertical error bars correspond to Poisson errors. 
}
  \label{Fig8}
\end{figure*}

Provided these features about CV segregation, the spatial distributions for NGC 6397, NGC 6752, 47 Tuc and $\omega$ Cen can be easily explained.   
Both core-collapsed clusters (NGC 6397 and NGC 6752) have very short $T_{\rm rel}$ ($\lesssim 0.7$ Gyr),
which makes the movement of their bright and faint CV population relatively fast. 
In addition, as their faint CVs have masses similar to MSTO stars, they 
are as segregated as MSTO stars.
On the other hand, the $T_{\rm rel}$ of 47 Tuc is longer ($\sim 3$ Gyr),
but according to Fig. \ref{Fig7}, still short enough to allow 
differences between bright and faint CVs, if their masses are different.
However, if the masses of CVs in 47 Tuc are similar in both bright and faint CVs,
as suggested by \citet{Rivera_2017}, even if its $T_{\rm rel}$ is short,
the faint CV population would still be as segregated as bright CVs and more 
centrally concentrated than MSTO stars, as shown in Fig. \ref{Fig7}. 
On the other hand, \citet{Henleywillis_2018} found that most of the CVs in $\omega$ Cen reside outside the central region, with only the two most luminous CVs lying deep inside the core. 
These results suggest that indeed the CVs in that cluster have not segregated towards the core given its large $T_{\rm rel}$.

In general, we expect that in GCs, both bright and faint CVs at the onset
of mass transfer are more centrally concentrated than MSTO stars, due
to their history during the course of the cluster evolution, not because
they are mostly dynamically formed in the core.

\subsection{Spatial Distribution: dynamically formed CVs}

We notice that we can also explain the spatial distribution of the
bright CVs that are dynamically formed close to the present-day,
in core-collapsed GCs.
If a pre-CV is dynamically formed in the core, it cannot stay 
in the core after the formation, as the process of forming 
a pre-CV dynamically is very energetic and such a pre-CV will be
expelled far from the core. However, if the cluster half-mass relaxation time
$T_{\rm rel}$ is short enough (as in core-collapsed GCs) to allow the pre-CV 
to segregate before it becomes a CV, such a population can come back to 
the central parts. Thus, forming a bright CV in the central parts.
If $T_{\rm rel}$ is relatively long, those bright CVs
that are dynamically formed will be found far from the central parts.

This process is illustrated in Fig.~\ref{FigR}, where we show the radial
position evolution for one dynamically formed CV (together with the host cluster
radii evolution).
Such a model is evolving towards collapse until about $0.5$ Gyr, when the 10 per cent black hole 
Lagrangian radius reaches its minimum. This is when the core
bounce \citep{Breen_2013} occurs and the black holes generate sufficient energy to support the
whole cluster evolution.
As such a cluster is supported by a subsystem of black holes, the energy generation
is relatively large, which causes the cluster to expand \citep{Breen_2013}, 
and the whole black hole subsystem stops to mass segregate at about 1.2 Gyr.
Such a feature leads to an increase in $T_{\rm rel}$ during the cluster evolution, 
and close to the present day, $T_{\rm rel}\approx3.5$ Gyr.

Regarding the CV, we notice that its progenitor (MS-MS binary) also initially segregates, 
as it is one of the most massive objects in the cluster. At the moment the core bounce occurs, 
the CV progenitor is violently ejected from the core in a four-body interaction with
another MS-MS binary. After that it takes around $5$ Gyr to segregate back to the
core. However, during this time, the binary starts unstable mass transfer and merges (due to the
high initial mass ratio) at $\approx1$ Gyr, leading to the formation of a single WD.
After the single WD returns to the core, it interacts with MS-MS binary and one of
such MS stars is replaced with the single WD, leading to a WD-MS binary at $\approx5.5$ Gyr.
Such a binary is dissolved in a three-body interaction with a single MS star at $\approx6.5$ Gyr.
After that, the single WD interacts again with a MS-MS binary at $\approx6.7$ Gyr, which
leads to the pre-CV formation when again a MS star is replaced with the WD intruder.
Notice that right after the pre-CV formation in the core, 
the pre-CV is kicked out from there, and takes a few Gyr ($\approx3$ Gyr) to segregate back to the inner parts.
That time is consistent with $T_{\rm rel}$. In this way, dynamically formed
CVs in cluster with shorter $T_{\rm rel}$ are expected to migrate back inwards faster. In other
words, the shorter $T_{\rm rel}$, the faster the dynamically formed CV segregates back
to the inner parts, after being ejected from the core due to the strong interaction.
This particular CV formation (i.e. onset of mass transfer) takes place close to the present day.

\subsection{Predicted number of detectable CVs and observed number of CVs and CV candidates}

After all this discussion about CV properties, we can now attempt to provide some useful
constraints for GC modelling. To do so, we show in Fig. \ref{Fig8}
the number of detectable CVs within $R_L$ against the cluster mass, including
the observational results.
We separate the models according to the IBP (Standard and Kroupa) and 
the CEP efficiency ($\alpha=$ 0.25, 0.50 and 1.00).

From Fig. \ref{Fig8}, it is clear that the number of
detectable CVs strongly depends on model assumptions, and clusters
with similar present-day masses can have very different numbers of
detectable CVs.
In any event, the maximum number of detectable CVs (for models clustering
around a particular mass) is still higher for greater GC masses, 
which is consistent with results for the entire
CV population (see Fig. \ref{Fig10}).
The fact that the number of detectable CVs depends on model
assumptions is quite convenient for us, since we can confront
them against observations and infer what are the assumptions that 
are likely to lead to consistent results.

With respect to the IBP, we can see that our 144 models set with the 
Standard IBP are unlikely to reproduce observations,
with the exception of M4.
They have always $\lesssim 5$ detectable CVs per cluster, irrespective
of initial conditions and stellar/binary evolution parameters, and they
have present-day masses roughly consistent with 
NGC 6397, NGC 6752 and M4.
On the other hand, models set with the Kroupa IBP are more likely to 
reproduce observational results for the four clusters considered here.
Indeed, when taking into account error bars and models with largest 
numbers of detectable CVs, and/or shortest $T_{\rm rel}$, the observed 
amount of CVs in 47 Tuc, NGC 6397, NGC 6752 and M4
are in a rough agreement with our 
results.
So, as we are not modelling
particular GCs, we can conclude that our results are roughly in agreement 
with observations, specially provided the small number statistics 
related to GCs with deep observations regarding CVs.

Regarding the CEP efficiency, we notice that models evolved with 
$\alpha=0.25$ and $0.5$ better reproduce the observed amounts of 
CVs.
This is not the case when $\alpha=1$ is considered, which leads
to smaller numbers of detectable CVs.
This result is consistent with the fact that the smaller the $\alpha$,
the greater the amount of CVs. 

It is interesting though that, our results suggest that the long-standing 
problem related to the deficit of CVs in GCs can potentially be solved, 
while carefully including in the modelling appropriate IBP, 
CV formation/evolution prescriptions and observational selection effects.
We also note that a conclusive answer for that is out of the scope of 
this work, as we would need many more models such that the GC initial parameter space would be better filled (see Table \ref{TabMODELS}).

Notice that $\omega$ Cen was not included in this analysis.
This is because its mass is much larger than the other GC masses and 
we do not have amongst our models any that massive.
However, we can extrapolate results presented in Section \ref{pdp} in order
to account for $\omega$ Cen. Assuming that best results are for the Kroupa
IBP, low CEP efficiency and considering that $\approx2-4$ per cent of
all CVs should be detectable (and only $\approx50$ per cent of them
are inside $R_L$), we predict that there should be $\approx50-250$ 
detectable CVs inside the half-light radius of $\omega$ Cen. 
This number is at least $\approx2$ times
higher than the observed amount of CVs in that cluster. 
In addition, a better optical characterization and membership determination of the unclassified X-ray sources in that cluster may increase the number of detected CVs. 
For example, \citet{Henleywillis_2018} extrapolated their results and estimated that there must be $\sim40\pm10$ CVs with $L_X \geq 10^{30}$ erg s$^{-1}$, but several less luminous CVs must be present.


\section{Discussion}
\label{discussion}

We have analysed a relatively large sample of 288 GC models, 
evolved with an up-to-date version of the \mocca code, 
with respect to IBPs and stellar/binary evolution prescriptions.
These models have a variety of different initial conditions spanning 
different values of the mass, size, King parameter, IBP, binary fraction, 
and Galactocentric distances. In addition, we also explored two parameters
of stellar/binary evolution, namely inclusion or not of fallback
and three different CEP efficiencies.

With respect to dynamical production and destruction of CV
progenitors, our results suggest that we should expect less CVs in dense GCs relative to the MW field, due to the fact that destruction of CV progenitors is more important in GCs than dynamical formation of CVs. 
Indeed, all our models have mass densities smaller than $\sim8.5\times10^{-5}$~CVs~M$\odot^{-1}$, which is smaller than the one in the Galactic field ($\sim10^{-4}$~CVs~per~M$_\odot$\footnote{Considering that the mid-plane CV space density in the
Milky Way is $\sim4\times10^{-6}$~pc$^{-3}$ \citep[][]{2012_pretorius} and that the mid-plane stellar mass density is $\sim4\times10^{-2}$~M$_\odot$~pc$^{-3}$ \citep[][]{Bovy_2017}.}). This is in agreement with results found by \citet{Cool_2013} for the massive cluster $\omega$ Cen. These authors determined a mass density of $\sim1.4\times10^{-5}$~CVs~per~M$_\odot$, which is consistent with our derived 
average density ($\sim2\times10^{-5}$~CVs~M$\odot^{-1}$), and about 1 order of magnitude smaller 
than the one in the Galactic field. We notice that this is also consistent with 47 Tuc,
which has a CV mass density of $\sim7\times10^{-5}$~CVs~M$_\odot^{-1}$, i.e. smaller than in the MW.
Finally, our results are also in agreement with recent analysis of
archival Chandra data by \citet{Cheng_2018}. These authors analysed
69 GCs and conclude that, different from what has been previously
thought, the weak X-ray populations, primarily
CVs and ABs, are under-abundant in GCs with respect to the 
Solar neighborhood and Local Group dwarf elliptical galaxies.

In addition, our results are also in agreement with the concentration of CVs found in $\omega$ Cen. 
\citet{Cool_2013} and \citet{Henleywillis_2018} found that there is no
much difference in the radial distribution between bright and faint CVs, which is in line with our Fig. \ref{Fig6}, since
$\omega$ Cen has a very long $T_{\rm rel}$.
We stress that the radial distribution of CVs in GCs is not indicative of their
formation channel, as suggested previously \citep[e.g.][]{Davies_1997}, but rather a 
consequence of the GC properties, mainly $T_{\rm rel}$, and the CV masses.

We found that, on average, more than half of the entire population of 
detectable CVs are outside the half-light radius, and future observations 
could aim to search for X-ray sources in regions not close to the cluster 
centres.
However, we notice that this should be considered an upper limit, as the 
real fraction of CVs outside the half-light radius varies from cluster to 
cluster and depends mainly on its $T_{\rm rel}$. 
Indeed, in Fig. \ref{Fig6} and \ref{Fig7} we show that
 (i) the greater the CV mass, the stronger the effect of mass segregation and, 
(ii) the shorter the cluster's $T_{\rm rel}$, the stronger the effect of mass segregation.
Said that, we expect that clusters with relatively long $T_{\rm rel}$ 
(longer than a few Gyr) will have a considerable amount of CVs outside their
half-light radius and the opposite for cluster with short $T_{\rm rel}$ (shorter than a few hundreds of Myr).
Our results then suggest that in clusters where $T_{\rm rel}$ are relatively long, we would be able to roughly double the number 
of GC CV candidates while looking for them outside $R_L$.
This would also allow us to obtain accurate spectra since the crowding 
should not be a big problem in the regions far away from the GC centres.

\begin{figure}
   \begin{center}
    \includegraphics[width=0.95\linewidth]{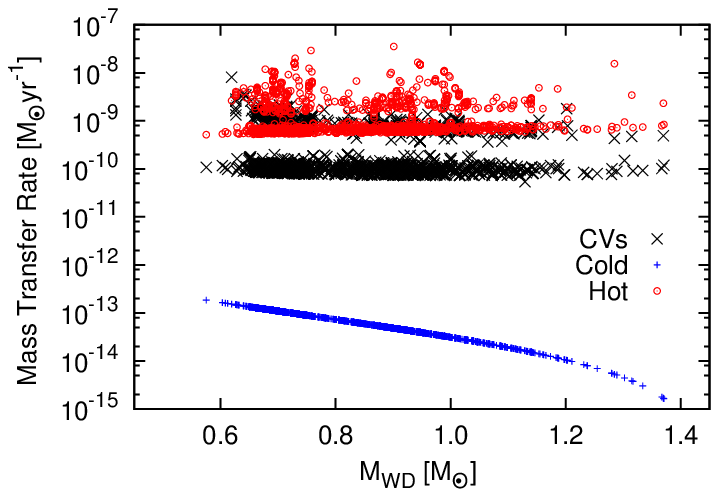}
    \includegraphics[width=0.95\linewidth]{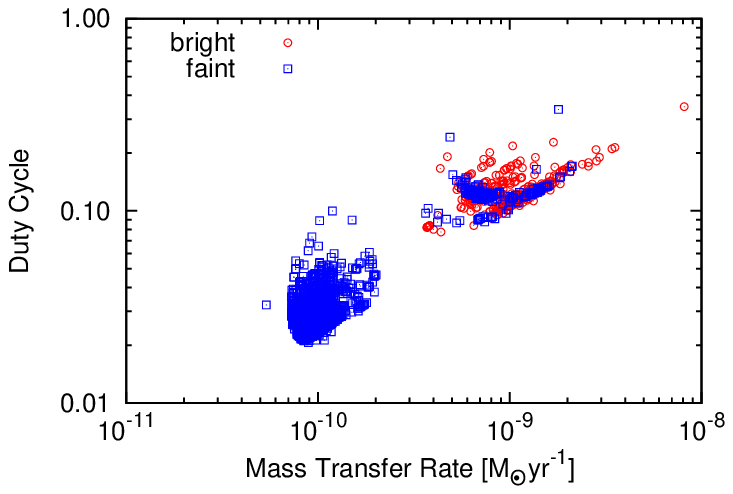}
    \end{center}
  \caption{Top panel: comparison of the values of the mass transfer rate required 
for a full CV disc to be globally hot and stable or globally cold and stable \citep[computed as described in][]{Belloni_2016a}
with respect to the mass transfer rate of detectable CVs. 
Bottom panel: Duty cycle \citep[computed with the empirical relation provided in][]{Britt_2015}
versus the mass transfer rate for the detectable CVs in the simulation.}
  \label{Fig9}
\end{figure}

One important long-standing question related to GC CVs is whether
they are predominately magnetic or not. In order to test the
hypothesis that most CVs in GCs are actually dwarf novae (DNe) with 
very short duty cycles, we first separate all detectable CVs according
to the stability in the accretion disc and compute their duty cycles.
In the top panel of Fig. \ref{Fig9} we show the values of the mass 
transfer rate required for a full CV disc to be globally hot and 
stable or globally cold and stable \citep[computed as described in][]{Belloni_2016a}
with respect to the mass transfer rate of detectable CVs. 
It clearly shows that more than 99 per cent of all detectable CVs
in our 288 models are DNe (black points between blue and red points). 
Additionally, the figure illustrates that for a great deal of them,
the instability takes place far away from the WD surface, since the detectable CV mass transfer rates (black points) are close to the
mass transfer rates required for having the disc globally hot (red points).
This indicates that the strength of the WD magnetic field 
has to be relatively strong in order to prevent the disc instability 
and, in turn, the outbursts in such systems.  Indeed, based on equation 8 in \citet{Ivanova_2006}, we computed the minimum magnetic field required to prevent DN outbursts 
for all detectable CVs in our simulations, 
and found that most of them would require a magnetic field stronger than
$\sim~3\times10^6$ G, which is consistent with magnetic fields in MW magnetic CVs.

In the bottom panel of Fig. \ref{Fig9} we depict their duty cycles
against their mass transfer rates, where the duty cycles were computed 
based on the empirical relation provided in \citet{Britt_2015}.
Note that most faint CVs have extremely short duty cycles ($\lesssim$ 6 per cent) and most
bright CVs have duty cycles greater than 10 per cent.
This then suggests that detecting outbursts amongst faint CVs is rather improbable.
On the other hand, part of the bright DNe would be recovered in multi-epoch
searches, such as those performed by \citet{Shara_1996} and \citet{Pietrukowicz_2008}.
Indeed, the completeness of their searches were computed based on DNe with
duty cycles consistent with the values we estimate for our bright CVs.
Now, provided that only $\approx 2-4$ per cent of all CVs would be detectable and
given that, on average, only $\approx 17$ per cent of detectable CVs are bright,
the number of DN outbursts found by these authors seem quite consistent with
our results. For example, in 47 Tuc, the number of bright CVs is $\sim 10$ 
(based on the NUV CMD), which implies that only around 3--4 DNe would be detected by
\citet{Shara_1996} (given their assumed duty cycles), in good agreement 
with the amount of DN outbursts recovered in 47 Tuc to date \citep{Shara_1996,Wilde_2015}.
Therefore, our results do not offer evidences for an overabundance of
magnetic CVs in GCs, in comparison with the fraction found in the MW.

An interesting very recent observational fact is that there is no significant
correlation between the number of bright X-ray CVs
and the cluster stellar encounter
rate \citep{Cheng_2018}, which is different from what has been
thought previously \citep[e.g.][]{Pooley_2003,Pooley_2006}. 
This indicates that bright CVs are a mix of 
dynamically formed and primordial CVs, and formation via typical
CEP might play a significant role.
We found here that, in general, most bright CVs come from primordial
CV progenitors, which is in agreement with the recent observational
result by \citet{Cheng_2018}.
In any event, we stress that the X-ray source population above 
$\sim10^{31}$ erg s$^{-1}$ includes different types of close binaries, 
such as chromospherically active stars
(which for example dominates the X-ray population in 47 Tuc \citep{2017Bhattacharya}),
millisecond pulsars, low-mass X-ray binaries, foreground and background objects, etc.
Finally, as pointed out by \citet{Cheng_2018},
taking the number of X-ray sources versus
the stellar encounter rate relation as solid evidence 
for a dynamical origin of the X-ray populations may be 
potentially an over-simplification of the issue.

One important observational result is that the number of bright CVs per cluster mass
in core-collapsed clusters is so far much higher than in non-core-collapsed clusters
\citep{Cohn_2010,Cool_2013,Lugger_2017,Rivera_2017,Henleywillis_2018}.
We found here that, on average, 
for non-core-collapsed models set with the Kroupa IBP,
$\sim 5-30$ per cent of the detectable CVs are bright,
which is consistent with 47 Tuc and $\omega$ Cen. 
Regarding core-collapsed Kroupa models, we found that
fraction to be $\sim5-45$ per cent. However, our core-collapsed
clusters with shortest $T_{\rm rel}$ usually have fractions higher
than $\sim 50$ per cent. We conclude then that our results
are also consistent with respect to NGC 6397 and NGC 6752, which are
core-collapsed GCs with very short $T_{\rm rel}$ and have
observed fractions of bright CVs in the range of $\sim40-60$ per cent.
Our results suggests then that the formation of CVs 
is indeed slightly favored through strong dynamical interactions in core-collapsed GCs, due to 
the high stellar densities in their cores.
However, selection effects might play an important role. Given the stellar crowding, the detection of CVs in GCs is X-ray biased compared to the field population. Additionally, many faint X-ray CV candidates have been detected only through ultraviolet observations instead of optical, 
as in the case of 47 Tuc \citep{Rivera_2017}, where 22 systems were detected for the first time using that technique. 

Our results suggest that models set with the Kroupa
IBP and low CEP efficiency better reproduce the amount of 
observed CVs in NGC 6397, NGC 6752 and 47 Tuc.
Low CEP efficiency is consistent with recent investigations that have concluded that
WD-MS binaries experience a strong orbital shrinkage during the CEP 
\citep[e.g.][]{Zorotovic_2010,Toonen_2013,Camacho_2014,Cojocaru_2017}.
The Kroupa IBP has successfully explained the observational features of young clusters,
stellar associations, and even binaries in old GCs
\citep[e.g.][and references therein]{Kroupa_2011,Marks_2012,Leigh_2015,Belloni_2017c,Belloni_2018a},
and our results provide an additional support to this IBP.

Although some of our models produce reasonable amounts of detectable CVs, 
we stress that better comparisons depend on better characterization of the GC CV candidates. 
Indeed, only a few GC CVs are spectroscopically confirmed so far. 
Thus it is an urgent demand to obtain more properties of the CV candidates, 
for example, the orbital periods and mass ratios.

With respect to results for the Standard IBP, we stress that
it always leads to $\lesssim 5$ detectable CVs per GC.
Then, albeit hard to imagine, if most CV candidates in the 
clusters discussed here does not turn out to be real CVs, 
then the Standard IBP would lead to reasonable
amounts.
In addition, the amount of predicted CVs might depend on the
assumed binary fraction while adopting the Standard IBP.
This is because by increasing the amount of binaries, we also
increase the amount of potential CV progenitors.
Said that, future simulations could test the influence of
the binary fraction on our results, since, if confirmed
that the Standard IBP cannot reproduce observed GC CV properties,
this brings an even stronger evidence supporting the Kroupa IBP.

Another fact to be considered here is that in spite of the fact that we
can easily apply any detection criteria for our simulated data, the same
is not true with respect to observations. Indeed, for instance, the
limiting $L_X$ luminosity and the region inside the cluster are not the 
same in all observed GCs considered here \citep{Cohn_2010,Lugger_2017,Cool_2013,Rivera_2017,Henleywillis_2018}. 
However, with our selected criteria, many of the obtained CV properties
from the simulated clusters are in general agreement with 
the current observations.

Finally, we stress that our comparisons involved as main parameter the
cluster mass. In this way, we cannot freely claim that
our models are fully suitable for the observed GCs discussed
here. In order to find best-fitting models for particular GCs, a
more elaborated approach is needed and more predicted GC properties 
need to be confronted with observations, such as surface brightness 
profile, velocity dispersion profile, local luminosity function, 
mass function, pulsar accelerations, etc. 
\citep[e.g.][]{Giersz_2009,Giersz_2011}. 
For this reason, our results should be interpreted as general 
statistical conclusions, rather than as an attempt to model particular GCs.


\section{Conclusions}
\label{conclusions}

Our main results can be summarised as follows.

(i) We found a strong correlation at a significant level between the fraction of destroyed
primordial CV progenitors and the initial stellar encounter rate, i.e. we found
that the greater the initial stellar encounter rate, the stronger the role of 
dynamical interactions in destroying primordial CV progenitors.

(ii) We show that dynamical destruction of primordial CV progenitors
is much stronger in GCs than dynamical formation of CVs.

(iii) We confirm that strong dynamical interactions are able to trigger
CV formation in binaries that otherwise would never become CVs, 
by expanding the primordial CV progenitor parameter space.

(iv) Different from what we found previously, here we find that dynamically
formed CVs and CVs formed under no/weak influence of dynamics have similar
WD mass distributions.

(v) We find that the detectable CV population is predominantly composed
of CVs formed via typical CEP ($\gtrsim 70$ per cent).
In addition, on average, only 
$\approx 2-4$  per cent of all CVs in a GC is likely to be detectable.

(vi) Even though amongst detectable CVs the fractions of bright/faint CVs and
CVs inside/outside the half-light radius change from model to model, we show that
the longer the cluster half-mass relaxation time, the higher the fraction of CVs that are outside the half-light radius (with upper limit of $\sim50$ per cent) and, 
on average, non-core-collapsed models tend to have small fractions of bright CVs, while core-collapsed models have higher fractions.

(vii) We show that the properties of bright and faint CVs can be understood by
means of the WD-MS and CV formation rates, their properties at their formation
times and cluster relaxation times. In this way, most detectable CVs have their 
WD-MS binaries formed before $\sim1$ Gyr and they take $\gtrsim 9$ Gyr to become CVs. 
This allows them to have enough time to sink to the central parts 
(being hard detached WD-MS binaries and having associated probability of being destroyed extremely small).
The fact that bright CVs are younger and more massive than faint CVs makes them, 
in general, more centrally concentrated, as observed in NGC 6397 and NGC 6752,
unless bright and faint CVs have similar total masses so that they will have similar
levels of mass segregation, as seen in 47 Tuc.

(viii) Even though we found the total number of CVs correlates with the host GC masses,
we also found that the number of detectable CVs is very sensitive to
model assumptions and that GC models with similar mass might have very different
numbers of detectable CVs.

(ix) Even though we had no intention of modelling particular GCs,
by comparing our results with observations, we show that,
amongst our models, those following
the Kroupa IBP and set with low CEP efficiency ($\lesssim 0.5$) better reproduce 
the observed amount of CVs and CV candidates in NGC 6397, NGC 6752, M4 and 47 Tuc.

(x)
Finally, we suggest that, in order to progress further with comparisons,
it is crucial to derive properties such as orbital period, mass transfer 
rate and mass ratio for the CV candidates.

\section*{Acknowledgements}

We would like to thank an anonymous referee for the comments and 
suggestions that helped to improve this manuscript.
DB was supported by the grants \#2017/14289-3 and \#2013/26258-4, S\~ao Paulo Research Foundation (FAPESP), and acknowledges partial support from the National Science Centre, Poland, through the grant UMO-2016/21/N/ST9/02938.
MG acknowledges partial support from National Science Center, Poland, through the grant UMO-2016/23/B/ST9/02732.
AA is supported by the Carl Tryggers Foundation for Scientific Research through the grant CTS 17:113.

\bibliographystyle{mnras}
\bibliography{references}

\bsp

\label{lastpage}

\end{document}